\documentclass[prd,nofootinbib,english]{revtex4-2}
\usepackage[utf8]{inputenc}

\usepackage{graphicx,color}
\usepackage{xcolor}
\usepackage{rotating}
\usepackage{amssymb}
\usepackage{amsfonts}
\usepackage{amsmath}
\usepackage{xspace}
\usepackage{mathtools}
\usepackage{verbatim}
\usepackage{comment}
\usepackage{soul}

\usepackage{physics}
\usepackage{graphicx}
\usepackage{hyperref}

\usepackage{accents}

\newtheorem{thm}{Theorem}

\newcommand\ie{\mbox{\textit{i.\,e.}}\xspace}

\newcommand\D{\mathrm{d}}

\newcommand{\Bf}{\bar{f}}

\newcommand{\ord}[2]{\accentset{(#2)}{#1}}

\begin{document}
\title{Weak equivalence principle and nonrelativistic limit of general dispersion relations}

\author{Manuel Hohmann}
\email{manuel.hohmann@ut.ee}
\affiliation{Laboratory of Theoretical Physics, Institute of Physics,  University of Tartu, W. Ostwaldi 1, 50411 Tartu, Estonia}

\author{Christian Pfeifer}
\email{christian.pfeifer@zarm.uni-bremen.de}
\affiliation{ZARM, University of Bremen, 28359 Bremen, Germany.}

\author{Fabian Wagner}
\affiliation{Dipartimento di Ingegneria Industriale, Universit\`a degli Studi di Salerno, Via Giovanni Paolo II, 132 I-84084 Fisciano (SA), Italy}
\affiliation{Institute of Physics, University of Szczecin, Wielkopolska 15, 70-451 Szczecin, Poland}

\begin{abstract}
We study the weak equivalence principle in the context of modified dispersion relations, a prevalent approach to quantum gravity phenomenology. We find that generic modified dispersion relations violate the weak equivalence principle. The acceleration in general depends on the mass of the test body, unless the Hamiltonian is either two-homogeneous in the test particles' 4-momenta or the corresponding Lagrangian differs from the homogeneous case by a total derivative only. The key ingredient of this calculation is a $3+1$ decomposition of the parametrization invariant relativistic test particle action derived from the dispersion relation. Additionally, we apply a perturbative expansion in the test particle's spatial velocity and the inverse speed of light.  To quantify our result, we provide a general formula for the E\"otv\'os factor of modified dispersion relations. As a specific example, we study the point-particle motion determined from the $\kappa$-Poincar\'e dispersion relation in the bicrossproduct basis. Comparing the ensuing non-vanishing E\"otv\'os factor to recent data from the MICROSCOPE experiment, we obtain a bound of the model parameter $\hat\Xi^{-1}\geq10^{15}{\rm GeV}/c^2$.
\end{abstract}

\maketitle

\section{Introduction}
The weak equivalence principle (WEP) is a cornerstone of classical gravitational physics. In its simplest form it can be stated as the fact that inertial mass and gravitational mass are at least proportional to each other.\footnote{We will not distinguish between active and passive gravitational mass, which one could also distinguish. This equality has  recently been verified to high accuracy \cite{Singh:2022wyp}.}  A more precise formulation reads, following C.M. Will \cite{Will:2014kxa},
\begin{quote}
    "The trajectory of a freely falling “test” body (one not acted upon by such forces as electromagnetism and too small to be affected by tidal gravitational forces) is independent of its internal structure and composition."
\end{quote}
One simple consequence of this statement is that two distinct and sufficiently small bodies (having possibly different masses and different elementary particle content) in a gravitational field are subject to the same gravitational acceleration. In other words, free fall is universal.

Possible violations of the WEP are conventionally quantified by means of the E\"otv\'os factor, which is the normalized difference of the spatial gravitational accelerations of two bodies with equal initial state of motion
\begin{align}
    \eta = 2 \frac{|\vec{a}_1 - \vec{a}_2|}{|\vec{a}_1 + \vec{a}_2|}\,.
\end{align}
By measuring the acceleration of two classical bodies (one made of titanium one made of platinum) in free fall, the validity of the WEP has just recently been confirmed to very high accuracy by the microscope mission \cite{MICROSCOPE:2022doy}, which has constrained the E\"otv\'os factor to $\eta \leq 10^{-15}$.

Thus, for classical matter and classical gravity there are currently no observational traces of deviations from the WEP. It is still an open question whether the WEP is valid for quantum matter \cite{Altschul:2014lua,Emelyanov:2021auc,Chiba:2022fzo,Blasone:2023yqz,Balsells:2023jdn}, in the context of quantum gravity \cite{Giacomini:2021aof,Bose:2022czr}, and how it should be best formulated or extended to a fundamental quantum-physics description of all interactions \cite{Zych:2015fka,Rosi:2017ieh,Padmanabhan:2020azn,Paunkovic:2022flx}. Does the classical WEP suffice, or is there the need for a quantum equivalence principle \cite{Zych:2015fka,Hardy:2019cef,Das:2023cfu}?

A necessary condition for the WEP to be implemented in a theoretical model of classical or quantum gravity and its effect on classical or quantum matter is that the spatial acceleration equation of point-like test particles or point-particle excitations of quantum fields is independent of the mass of the particle. If this is not the case, the acceleration of the test particles of different masses differs. Therefore, a deviation from the WEP, in form of a non-vanishing E\"otv\'os factor, ensues. This feature has been claimed to emerge from non-Lorentz invariant couplings between matter in gravitational fields in the standard model extension \cite{Tasson:2016xib}, from rainbow gravity \cite{Ali:2014aba}, generalized uncertainty principles \cite{Ali:2011ap,Gao:2017zch,Casadio:2020rsj,Tkachuk:2012gyq,Gnatenko:2022ttc}, $\kappa$-deformed quantum mechanics \cite{Wagner:2023fmb} or any non-universal coupling between gravity and matter.\vspace{6pt}

In this article we classify the validity of the WEP in the context of modified dispersion relations (MDRs) arising from violations of or deviations from local Lorentz invariance in the context of quantum gravity phenomenology. The underlying assumption of our study is that the modified dispersion relation determines the motion of point particles via the Hamilton equations of motion and that no further input is needed.\vspace{6pt}

Since a self-consistent fundamental theory of quantum gravity is still elusive, phenomenological models are employed to effectively describe aspects of the interaction of matter with quantum gravity \cite{Addazi:2021xuf}. The propagation of particles and fields on several models of quantum spacetime, such as non-commutative spacetimes, string theory backgrounds or spacetimes with a minimal length, are for example described by modified dispersion relations (MDRs). They are determined by a Hamilton function whose Hamilton equations of motion determine the worldlines of point particles that are used to probe the spacetime geometry, \ie to probe gravity. Deviations from the classical geodesics of a metric solving the Einstein equations in general relativity can then be investigated if they can be interpreted as quantum-gravity effects on the particle trajectories.

We study the relation between MDRs and the WEP, by deriving the spatial acceleration of test particles that are subject to the MDR. We find that for generic MDRs the spatial acceleration is mass dependent, and thus many classes of MDRs violate the WEP. However, we also identify a necessary and sufficient condition on the modification of the generally relativistic dispersion relations such that the WEP is valid, and we identify how MDRs satisfying the WEP can be constructed.

To reach this result we need to determine as key ingredient the spatial acceleration equation from MDRs. In general relativity, this is done by $3+1$ decomposing the geodesic equation on spacetime into a spatial and temporal part, and by using the time coordinate as parameter for the particle wordlines. Then, a $1/c$ expansion allows to study the acceleration of initially slowly moving test particles, seen by observers whose proper time coincides with the chosen coordinate time, to any desired post-Newtonian (pN) order.

Here, we follow an analogous procedure, just that we cannot use metric spacetime geometry, as in general relativity, to identify and derive the spatial acceleration. It turns out that Finsler geometry is the perfect tool to use instead \cite{Girelli:2006fw,Pfeifer:2019}. In Finsler geometry the geometry of spacetime is derived from a general length measure for curves, physically speaking from a general geometric proper time clock, that is not necessarily derived from a metric. In short, Finsler geometry is a straightforward generalisation of metric, Riemannian, geometry "without the quadratic restriction" \cite{Chern}, \ie not using a root of a quadratic length measure for vectors, but a general norm. We then present a $3+1$ decomposition and a $1/c$ pN expansion for a general Finslerian length measure for the first time. This important intermediate result lays the foundation for the pN analysis of Finsler gravity theories \cite{Hohmann:2019sni} in the future. The spatial acceleration equation of MDRs is then obtained by specifying the general Finsler length to the one induced by MDRs, from which we obtain the E\"otv\'os factor and the conditions under which it vanishes and the WEP is valid.

The structure of the article is as follows. In Section~\ref{sec:FromMDRsToFinsler} we recall how dispersion relations can be understood in terms of Hamilton functions on the 1-particle phase space of spacetimes and how they lead to a Finslerian spacetime geometry, describing the motion of point particles through the Helmholtz action. Afterwards, in Section~\ref{sec:NR} we $3+1$ decompose general Finslerian length measures for curves and perform a pN expansion before we derive the general spatial acceleration equation from the Finsler geodesic equation in Section~\ref{sec:acceq}. Finally, in Section~\ref{sec:EP} we apply the general findings to the Finsler length measure that is induced by an MDR an find in Theorem~\ref{thm:WEPMDR} that the modification of general relativistic dispersion relation must be $2$-homogeneous in the momenta, or the coupling constant must be mass dependent, in order to preserve the validity of the WEP for particles subject to the MDR. Also, we quantify violation of the WEP by deriving the E\"otv\'os factor for MDRs, specifically a class often employed in $\kappa$-Minkowski models, and we propose a type of MDR that satsifies the WEP by construction. Finally we close the article with a discussion and conclusion in Section~\ref{sec:Discussion}.

The notational conventions used in this article are such that the metric signature is $(+,-,-,-)$, indices $a,b,c,...$ run from $0,..,3$ while indices $\alpha, \beta, ...$ run from $1,..,3$.

\section{Finsler geometry and particle motion from dispersion relations}\label{sec:FromMDRsToFinsler}
Modified dispersion relations are employed to describe the propagation of particles and fields on quantum spacetime effectively \cite{Addazi:2021xuf}. In this section we briefly recall the mathematical description of dispersion relations as Hamilton functions and how they define a Finslerian spacetime geometry. Also, we discuss the main aspects of Finsler geometry that are used in this article. In particular we recall and point out, that the trajectories of free point particles subject to dispersion relations can be understood as Finsler geodesics. The latter will be the key ingredient to analyse the WEP in the context of MDRs in section \ref{sec:EP}.

Mathematically, dispersion relations are encoded in Hamilton functions $H(x,p)$ on the $1$-particle phase space of spacetime $S$, which is given by the cotangent bundle $T^*S$ of $S$ \cite{Barcaroli:2015xda}. Here, the variables $(x,p)$ denote the position $x^a$ of a particle and its $4$-momentum components $p_a$ understood as the components of the momentum covector $P=p_a dx^a\in T^*_xS$.

The Hamilton functions determine the dispersion relation and the particle motion via their level sets and the Hamilton equations of motion
\begin{align}
    H(x,p) = M^2\,,\quad \dot p_a + \partial_a H = 0\,,\quad \dot x^a = \bar\partial^a H\,,
\end{align}
where $M$ is the invariant mass parameter, and we used the abbreviations
\begin{equation}
\partial_a = \frac{\partial}{\partial x^a}\,, \quad
\bar{\partial}^a = \frac{\partial}{\partial p_a}\,,
\end{equation}
for the partial derivatives.

To fully understand the impact of MDRs on particle motion it is convenient to change from the Hamiltonian phase space description in terms of positions $x$ and momenta $p$ to the corresponding spacetime description in terms of positions $x$ and velocities $\dot x$. The main tool required to do so is the Helmholtz action of free massive point particles subject to a dispersion relation
\begin{align}\label{eq:HAc}
    S_H[x,p,\lambda] = \int d\tau (p_a \dot x^a - \lambda f(H(x,p),M))\,.
\end{align}
Variation with respect to $\lambda$ imposes the dispersion relation, where $f(H(x,p),M)$ is chosen such that $f(H(x,p),M) = 0$ implies $H(x,p)=M^2$ and variation with respect to $p$ allows to express $p$ in terms of $\dot x$ and $\lambda$. Inserting both of these expressions into the Helmholtz action, we can rewrite it as \cite{Girelli:2006fw,Lobo:2020qoa}
\begin{align}\label{eq:Fins}
    S[x] = S_H[x, p(x,\dot x, \lambda(x,\dot x)),\lambda(x,\dot x)] = M c \int d\tau F(x,\dot x)\,.
\end{align}
It turns out that $F(x,\dot x)$ is a so called Finsler function, which determines the effective Finslerian geometry of the quantum spacetime, see Appendix \ref{appx:Helmholtz} for details on how to derive $F(x,\dot x)$. One of its most important properties is that it is $1$-homogeneous with respect to $\dot x$
\begin{align}\label{eq:Fhom}
    F(x,\sigma\dot x) = \sigma F(x,\dot x)\quad  \forall \sigma >0\,,
\end{align}
which makes the action $S[x]$ parametrization invariant. This feature allows us to interpret the action $S$ as a geometric length measure for curves on spacetime, and, for timelike curves, as the proper time functional. The time which passes between two events $A$ and $B,$ which are joined by a timelike curve $\gamma = (x(\tau))$ with $\gamma(\tau_A)=A$ and $\gamma(\tau_B)=B$, is then given by
\begin{align}\label{eq:propt}
    \Delta\tau_{AB} = \frac{1}{c} \int_{\tau_A}^{\tau_B} d\tau \ F(x,\dot x)\,.
\end{align}
Point-particle motion is determined by $L=F^2$, which serves as free relativistic point-particle Lagrangian, and leads to the Finsler geodesic equation as Euler Lagrange equations of \eqref{eq:Fins}
\begin{align}\label{eq:geod}
    \frac{d}{d\tau} \dot\partial_a L - \partial_a L = 0
    \ \Leftrightarrow \
    \ddot x^a + 2 G^a(x,\dot x) = 0\,,
\end{align}
as point-particle equation of motion. The so called spray-coefficients $G^a$ generalize the Levi-Civita connection from usual Lorentzian spacetime geometry and are uniquely determine by $L$ and its derivatives
\begin{align}\label{eq:geodspray}
G^a(x,\dot x) = \frac{1}{4} g^L{}^{ab}\left( \dot x^c \partial_c \dot \partial_b L - \partial_b L\right)\,,
\end{align}
where
\begin{equation}
\dot{\partial}_a = \frac{\partial}{\partial\dot{x}^a}\,.
\end{equation}
Here $g^L{}^{ab}$ denotes the inverse of the Finsler metric
\begin{align}\label{eq:finsmet}
    g^L_{ab}(x,\dot x) = \frac{1}{2} \dot\partial_a  \dot\partial_b L\,,
\end{align}
which can be thought of as direction dependent, but scale invariant spacetime metric. For more details on Finsler spacetime geometry and its application to gravitational physics beyond MDRs see \cite{Hohmann:2021zbt,Hohmann:2020yia,Hohmann:2019sni,Pfeifer:2019} and references therein.

Since $F$ is derived from the Hamiltonian $H$ for particles which satisfy $H(x,p)=M^2$, in general $F$, and thus also the proper time, will depend on the mass parameter $M$\footnote{This the concept of an ideal clock turns into a 1-parameter family of ideal clocks, parametrized by the particles' mass.} in a non-trivial way as we will explicitly see in section \ref{sec:EP} for modified dispersion relations. There, we determine conditions on $H$ such that the spatial acceleration equation, which we derive from the Finsler geodesic equation in section \ref{sec:acceq}, does not depend on the mass of the particles, and thus the WEP is valid.

The Finsler function $F(x,\dot x)$ defines the free relativistic point-particle motion from the action \eqref{eq:Fins}. In order to analyse the WEP properly already at the nonrelativistic level, we next derive the generic $3+1$ decomposition and the nonrelativistic limit of general Finsler functions.

\section{The nonrelativistic expansion of the point-particle action}\label{sec:NR}
To analyse the $3+1$ decomposition and the nonrelativistic limit of a Finslerian geometry, let us fix the coordinate system $x^a = (ct, x^\alpha)$ and define
\begin{align}
    \epsilon^\alpha = \frac{\dot x^\alpha}{c \dot t} = \frac{v^\alpha}{c}\,.
\end{align}
Then, we can rewrite the Finslerian free point-particle action \eqref{eq:Fins}, where we display a factor of $c$ explicitly to match the usual units of an action, as
\begin{align}\label{eq:ppact}
    S[x] = M c \int d\lambda\ F(ct, x^\alpha, c \dot t, \dot x^\alpha)
    = M c^2 \int dt\ F(ct, x^\alpha, 1, \epsilon^\alpha)
    \equiv M c^2 \int dt\ f(ct, x^\alpha, \epsilon^\alpha)\,.
\end{align}
Here we used the one-homogeneity of $F$ in the velocities \eqref{eq:Fhom} and introduced the dimensionless function $f(ct, x^\alpha, \epsilon^\alpha)$. So far, this is just a rewriting and does not take into account whether the considered particles are slow nonrelativistic or fast relativistic.

To introduce a velocity order, we expand $f$ in  a power series of the small velocity parameters $\epsilon^\alpha$
\begin{align}
 f\left(ct,x^\alpha,\epsilon^\alpha\right)
 = \sum_{I=0}^{\infty}\frac{1}{I!}  \Bf_{\alpha_1...\alpha_I} \epsilon^{\alpha_1}\dots\epsilon^{\alpha_I}\,,\label{eqn:LagExp}
\end{align}
where the coefficients
\begin{align}
\Bf_{\alpha_1...\alpha_I} \equiv \left. \frac{\partial}{\partial\epsilon^{\alpha_1}}\ldots \frac{\partial}{\partial\epsilon^{\alpha_I}}f(ct,x^\alpha,\epsilon^\alpha)\right|_{\epsilon^\alpha = 0}\,,
\end{align}
transform as tensors under foliation (the hypersufaces $t=\text{const}$) preserving diffeomorphisms.  In general, these fields themselves may implicitly depend on the speed of light beyond the dependence of $f$ through the coordinates. Sometimes this is hidden in constants, for example the Schwarzschild radius $r_s=\frac{2 GM}{c^2}$ appearing in the components of the Schwarzschild metric. We formally expand the coefficients as
\begin{equation}
    \Bf_{\alpha_1\dots\alpha_I}=\sum_{J=0}^\infty \frac{\ord{\Bf}{J}_{\alpha_1\dots\alpha_I}}{c^J},\label{eqn:ExtFieldExp}
\end{equation}
where the coefficients $\ord{\Bf}{I}_{\alpha_1\dots\alpha_I} = \ord{\Bf}{I}_{\alpha_1\dots\alpha_I}(t,x^\alpha)$ do not harbour any intrinsic dependence on $c$ anymore. These expansions imply that the point-particle action can be expressed as
\begin{equation}
    S[x] = M c^2 \int dt \sum_{J=0}^\infty\frac{1}{c^J}\sum_{I=0}^J\frac{1}{I!}\
    \ord{\Bf}{J-I}_{\alpha_1\dots\alpha_{I}}v^{\alpha_1}\dots v^{\alpha_{I}}.
\end{equation}
In the above sum, $I$ counts the number of derivatives, while $J-I$ counts the expansion order of the fields $\bar f_{\alpha_1...\alpha_I}$ in accordance with \eqref{eqn:ExtFieldExp}. Consequently, $J$ counts the order in $1/c$. We thus obtain the expansion of the action to order $c^{-2}$
\begin{align}
    S[x] = M \int \D t
    &\bigg[
    c^2 \ord{\Bf}{0}
    + c \left(\ord{\Bf}{1}+\ord{\Bf}{0}_\alpha v^\alpha\right)
    + \left(\ord{\Bf}{2}+\ord{\Bf}{1}_\alpha v^\alpha+ \tfrac{1}{2}\ord{\Bf}{0}_{\alpha\beta} v^\alpha v^{\beta}\right)\nonumber \\
    &+ \frac{1}{c}\left(\ord{\Bf}{3}+\ord{\Bf}{2}_\alpha v^\alpha+\tfrac{1}{2}\ord{\Bf}{1}_{\alpha\beta} v^\alpha v^{\beta}+\tfrac{1}{3!}\ord{\Bf}{0}_{\alpha\beta\gamma} v^\alpha v^{\beta} v^{\gamma}\right)\label{eqn:PertLag}\\
    &+ \frac{1}{c^2} \left( \ord{\Bf}{4}+\ord{\Bf}{3}_\alpha v^\alpha+\tfrac{1}{2}\ord{\Bf}{2}_{\alpha\beta} v^\alpha v^{\beta}+\tfrac{1}{3!}\ord{\Bf}{1}_{\alpha\beta\gamma} v^\alpha v^{\beta} v^{\gamma} + \tfrac{1}{4!} \ord{\Bf}{0}_{\alpha\beta\gamma\delta} v^\alpha v^{\beta} v^{\gamma} v^{\delta} \right)
     + \mathcal{O}(c^{-3})\bigg].\nonumber
\end{align}
It is at this point, \ie when we cut off the series, that we designate the coordinate time $t$ as the global time underlying the nonrelativistic expansion, with respect to which we assume particle motion to be slow. We expanded the Finsler function up to order $c^{-2}$ since, in the metric case, $F = \sqrt{g_{ab}(x)\dot x^a \dot x^b}$ and thus $f = \sqrt{g_{00} + 2 g_{0\alpha} \epsilon^\alpha + g_{\alpha\beta}\epsilon^\alpha \epsilon^\beta}$, this is where the first post-Newtonian corrections appear in the (p)pN-expansion \cite{Will:2018bme,Weinberg}. Conventionally, this expansion is predicated on the assumptions
\begin{align}\label{eq:normalPN}
    \ord{g}{0}_{00}&=1,& \ord{g}{1}_{00}&=0,& \ord{g}{2}_{00}&=2\phi,& \ord{g}{3}_{00}&=0,& \ord{g}{4}_{00}&\neq0,\nonumber\\
    \ord{g}{0}_{0\alpha}&=0,& \ord{g}{1}_{0\alpha}&=0,& \ord{g}{2}_{0\alpha}&=0,& \ord{g}{3}_{0\alpha}&\neq0,& \ord{g}{4}_{0\alpha}&=0,\\
    \ord{g}{0}_{\alpha\beta}&=-\delta_{\alpha\beta},& \ord{g}{1}_{\alpha\beta}&=0,& \ord{g}{2}_{\alpha\beta}&\neq0,& \ord{g}{3}_{0\alpha}&=0,& \ord{g}{4}_{\alpha\beta}&\neq0\,,\nonumber
\end{align}
and one finds that the $c^2$-term is constant, the $c^1$-contribution vanishes and the Newtonian $c^0$-term is given by
\begin{align}
    \ord{\Bf}{2}+ \tfrac{1}{2}\ord{\Bf}{0}_{\alpha\beta} v^\alpha v^{\beta}\,,\quad
    \ord{\Bf}{2} = \tfrac{1}{2} \ord{g}{2}_{00} = \phi\,,\quad
    \ord{\Bf}{0}_{\alpha\beta} = \ord{g}{0}_{\alpha\beta} = -\delta_{\alpha\beta}\,.
\end{align}
Furthermore, the $c^{-1}$-contribution vanishes, and the first non-trivial relativistic corrections appear at order $c^{-2}$
\begin{align}
    \ord{\Bf}{4} = \tfrac{1}{2} \ord{g}{4}_{00} - \tfrac{1}{8} \ord{g}{2}_{00}{}^2 \,,\quad
    \ord{\Bf}{3}_\alpha = \ord{g}{3}_{0\alpha}\,,\quad
    \ord{\Bf}{2}_{\alpha\beta} = \ord{g}{2}_{\alpha\beta} - \tfrac{1}{2} \ord{g}{2}_{00}\ \ord{g}{0}_{\alpha\beta} \,,\quad
    \ord{\Bf}{1}_{\alpha\beta\gamma} = 0\,,\quad
    \ord{\Bf}{0}_{\alpha\beta\gamma\delta} = - \tfrac{4!}{8} \ord{g}{0}_{(\alpha\beta} \ord{g}{0}_{\gamma\delta)}\,.
\end{align}
Since we seek to analyse the nonrelativistic (slow velocity) limit of relativistic Finsler functions in general, we will not impose such constraints a priori.

Next, we derive the spatial acceleration equation, solely expressed in terms of $f$, and identify necessary conditions so that a slow-velocity limit ($c\to\infty$) with well-defined particle motion exists. Afterwards we evaluate the acceleration equation for MDRs to investigate the validity of the WEP in Section \ref{sec:EP}.

\section{From the geodesic equation to the spatial acceleration equation}\label{sec:acceq}
To understand the point-particle mechanics of the nonrelativistic limit, we investigate the geodesic equation \eqref{eq:geod}. In terms of the $3+1$ coordinates  $(ct, x^\alpha)$ we obtain
\begin{align}
    \ddot x^0  + 2 G^0(ct, x^\alpha, c\dot t, \dot x^\alpha) &= 0 \Leftrightarrow c \ddot t + 2 G^0(ct, x^\alpha, c\dot t, \dot x^\alpha) = 0\,,\\
    \ddot x^\beta  + 2 G^\beta(ct, x^\alpha, c\dot t, \dot x^\alpha) &= 0\,,
\end{align}
which can be recombined to obtain the spatial acceleration
\begin{align}\label{eq:NON-RELacc}
    \frac{d^2 x^\alpha}{d t^2} = \frac{d}{dt}\left( \dot x^\alpha \dot t^{-1}\right)
    =  \frac{2 }{\dot t^2} \left( - G^\alpha(ct, x^\alpha, c \dot t, \dot x^\alpha) + G^0(ct, x^\alpha,  c \dot t, \dot x^\alpha) \epsilon^\alpha \right)\,.
\end{align}
Thus, the goal of this section is to express the geodesic spray coefficients $G^0$ and $G^\alpha$ in a power series in $c^{-1}$. We proceed in two steps: First, we expand all required quantities which are contained in the spray geodesic coefficients, \eqref{eq:geodspray}  in terms of $f$ from $L=F^2 = c^2 \dot t^2 f^2$ to derive the spatial acceleration equation \eqref{eq:acceqallorders}. Afterwards, we use \eqref{eqn:LagExp} to obtain the nonrelativistic expansion up to order $1/c^2$ in \eqref{eq:accnonrelsimp}.

As short-hand notation, we use $f=f(ct, x^\alpha, \epsilon^\alpha)$, $f_{\alpha} = \partial f/\partial \epsilon^\alpha$ and $f_{\alpha\beta} =  \partial^2 f/\partial \epsilon^\beta \partial \epsilon^\alpha$ in the following. Note that these fields differ from the coefficients introduced in \eqref{eqn:LagExp} because the latter, contrary to the former, are evaluated at $\epsilon^\alpha= 0.$

\subsection{The Finsler metric and its inverse}
We start with the partial derivatives $\dot \partial$ with respect to velocities to construct the Finsler metric and its inverse:
\begin{align}
    \dot \partial_0 L
    &= \frac{1}{c} \frac{\partial}{\partial \dot t} F^2
    = 2 c \dot t f \left( f - \epsilon^\alpha  f_\alpha\right)\,,\\
    \dot \partial_\alpha L
    &= \frac{\partial}{\partial \dot x^\alpha} F^2
    = 2 c \dot t f  f_\alpha \,.
\end{align}
The Finsler metric is then given by
\begin{align}
    \frac{1}{2} \dot \partial_0 \dot \partial_0 L = g_{00} &= -N^2 + h_{\gamma\delta} N^{\gamma} N^{\delta}
    =  \left( f  - \epsilon^\alpha  f_\alpha \right)^2 + f \epsilon^\alpha \epsilon^\beta f_{\alpha\beta} \,,\\
    \frac{1}{2} \dot \partial_0 \dot \partial_\alpha L = g_{0\alpha} &= h_{\alpha\beta}N^\beta
    = f  f_\alpha  - \epsilon^\beta  \left(f_\alpha f_\beta + f f_{\alpha\beta} \right)\,,\\
    \frac{1}{2} \dot \partial_\beta \dot \partial_\alpha L = g_{\alpha\beta} &= h_{\alpha\beta}
    = f f_{\alpha\beta} + f_\beta f_\alpha \,,
\end{align}
where we introduced the ADM variables, \ie the shift vector $N^\alpha$ and the lapse function $N$. The inverse $h^{\alpha\beta}$ of the spatial metric can be calculated to be
\begin{align}
    h^{\alpha\beta} = \frac{1}{f}\left(f^{\alpha\beta} - \frac{f^{\alpha\gamma} f_\gamma \ f^{\delta\beta}f_\delta}{f + f^{\sigma\rho}f_\rho f_\sigma } \right) \quad \Rightarrow \quad
    h^{\alpha\beta}f_\beta = \frac{f^{\alpha\beta}f_\beta}{f + f^{\sigma\rho}f_\rho f_\sigma}\,.
\end{align}
Here $f^{\sigma\rho}$ denotes the inverse of the matrix $f_{\sigma\rho}$. The inverse Finsler metric then reads
\begin{align}
    g^{00} &= -\frac{1}{N^2}
    = \frac{f + f^{\alpha\beta}f_\alpha f_\beta}{f^3}\\
    g^{0\alpha} &= \frac{N^\alpha}{N^2}
    = -  \frac{f^{\alpha\beta}f_\beta}{f^2}  + \frac{\epsilon^\alpha(f + f^{\sigma\rho}f_\sigma f_\rho)}{f^3}\\
    g^{\alpha\beta} &= h^{\alpha\beta} + \frac{N^\alpha N^\beta}{N^2}
    =\frac{1}{f}\left(f^{\alpha\beta} - \frac{\epsilon^\alpha f^{\gamma\beta}f_\gamma}{f} - \frac{\epsilon^\beta f^{\gamma\alpha}f_\gamma}{f} + \frac{\epsilon^\alpha \epsilon^\beta (f + f^{\sigma\rho}f_\sigma f_\rho)}{f^2} \right)\,.
\end{align}
With these tools at hand, we proceed to expand the geodesic spray coefficients.

\subsection{The Geodesic spray}
First, we express the geodesic spray in terms of $f$. Hence, we need to derive, see\eqref{eq:geodspray},
\begin{align}
    4 G^0(ct, x^\beta, c \dot t, \dot x^\beta)
    &= g^{00}\left(\dot x^0 \partial_0 \dot\partial_0 L + \dot x^\alpha  \partial_\alpha \dot \partial_0 L - \partial_0 L\right)
    + g^{0\alpha}\left(\dot x^0 \partial_0 \dot \partial_\alpha L + \dot x^\beta \partial_\beta \dot \partial_\alpha L
    - \partial_\alpha L \right)\,\\
    4 G^\alpha(ct, x^\beta, c \dot t, \dot x^\beta)
    &= g^{\alpha0}\left(\dot x^0 \partial_0 \dot\partial_0 L + \dot x^\alpha  \partial_\alpha \dot \partial_0 L - \partial_0 L\right)
    + g^{\alpha\beta}\left(\dot x^0 \partial_0 \dot \partial_\beta L + \dot x^\gamma \partial_\gamma \dot \partial_\beta L
    - \partial_\beta L \right)\,.
\end{align}
The brackets can easily be expanded in derivatives acting on $f$
\begin{align}
    \dot x^0 \partial_0 \dot\partial_0 L + \dot x^\alpha  \partial_\alpha \dot \partial_0 L - \partial_0 L
    &= 2 \dot t^2 f \left[ c \left(\partial_t f + 2 v^\alpha \partial_\alpha f\right)
    - \left( v^\alpha( f_\alpha \partial_t\ln f + \partial_t f_\alpha)
    + v^\alpha v^\beta ( f_\beta \partial_\alpha\ln f + \partial_\alpha f_\beta )\right)\right]\,,\\
    \dot x^0 \partial_0 \dot \partial_\alpha L + \dot x^\beta \partial_\beta \dot \partial_\alpha L
    - \partial_\alpha L
    &= 2 \dot t^2  f\left[ c \left( f_\alpha \partial_t\ln f + \partial_t f_\alpha + v^\beta (f_\alpha \partial_\beta \ln f + \partial_\beta f_\alpha)\right) - c^2 \partial_\alpha f\right] \,.
\end{align}
Combining these expressions with the components of the inverse metric leads to
\begin{align}
    \frac{2 G^0}{\dot t^2}
    &= c^2 \frac{ f^{\alpha\beta} f_{\beta}\partial_\alpha f}{f} + c \frac{\partial_t f + v^\alpha \partial_\alpha f - f^{\alpha\beta}f_\beta (\partial_t f_\alpha + v^\gamma \partial_\gamma f_\alpha)}{f}\,,\\
   \frac{2 G^\alpha}{\dot t^2}
   &= - c^2 f^{\alpha \beta } \partial_{\beta }f
   + c \left( \frac{v^{\alpha } f^{\beta\gamma } f_\gamma\partial_{\beta }f}{f}
   +  f^{\alpha \beta } (\partial_t f_{\beta } +  v^{\gamma } \partial_{\gamma }f_{\beta })\right)
   + \frac{v^{\alpha } ( \partial_t f + v^{\beta }\partial_{\beta }f - f^{\beta\gamma }f_\gamma (\partial_t f_{\beta } +  v^{\gamma } \partial_{\gamma }f_{\beta }))}{f}\,.
\end{align}
Given these results, we are finally in the position to obtain the nonrelativistic limit of the acceleration equation.

\subsection{The small-velocity limit of the acceleration equation}
Combining all terms we obtained in the previous two subsections in \eqref{eq:NON-RELacc}, we find the $3+1$ decomposed spatial acceleration equation, which is of the surprisingly simple form
\begin{align}\label{eq:acceqallorders}
    \boxed{
    \frac{d^2 x^\alpha}{d t^2} = f^{\alpha \beta } \left(c^2  \partial_{\beta }f - c (\partial_t f_\beta + v^\gamma \partial_\gamma f_\beta )\right)\,.}
\end{align}
This is the $3+1$ decomposition of any $1$-homogeneous point-particle Lagrangian whose square is $L = c^2 \dot t^2 f(c t, x^\alpha, v^\alpha c^{-1})^2$, to all orders in $1/c$. We stress that \eqref{eq:acceqallorders} applies to both relativistic and nonrelativistic motion, \ie no power series expansion in $c^{-1}$ has been performed yet.

The same equations of motion can be obtained from the Euler-Lagrange equations
\begin{align}\label{eq:geodspat}
    \frac{d}{dt} \frac{\partial f}{\partial v^\alpha} - \frac{\partial f}{\partial x^\alpha} = 0\quad
    \Leftrightarrow
    \frac{1}{c}\frac{d}{dt} \frac{\partial f}{\partial \epsilon^\alpha} - \frac{\partial f}{\partial x^\alpha} = 0\,,\ \epsilon^\alpha = \frac{v^\alpha}{c}\,.
\end{align}
In other words, we can interpret any constant multiple $\lambda$ of $f = f(ct,x^\alpha,v^\alpha c^{-1})$ as Lagrange function $\mathcal{L} = \lambda \mathcal{L}(t,x^\alpha,v^\alpha) = \lambda f(ct,x^\alpha,v^\alpha c^{-1})$ for $3+1$ decomposed point-particle motion, $\lambda \in \mathbb{R}$. In general $\lambda$ can dependent on $c$. We call $f$ the \emph{spatial Lagrangian}.

Further expanding $f$ in terms of $\epsilon^\alpha$ and $1/c$ in accordance with Eqs. \eqref{eqn:LagExp} and \eqref{eqn:ExtFieldExp}, we obtain the general slow-velocity expansion of the contravariant spatial acceleration,
\begin{align}\begin{split}
        f_{\alpha\beta} \frac{d^2 x^\alpha}{d t^2}
        &= \left(\sum_{I=0}^\infty \frac{1}{I!} \left(c^2  \partial_{\beta }\bar f_{\alpha_1...\alpha_I} - c (\partial_t \bar f_{\beta\alpha_1...\alpha_I} + v^\gamma \partial_\gamma \bar f_{\beta\alpha_1...\alpha_I} )\right)\epsilon^{\alpha_1}...\epsilon^{\alpha_I}\right)\\
        &=  \sum_{I=0}^\infty \frac{1}{I!}
        \left( \sum_{K=0}^\infty \frac{1}{c^{K-2+I}} \partial_{\beta }\ord{\bar f}{K}_{\alpha_1...\alpha_I}
        - \frac{1}{c^{K-1+I}} (\partial_t \ord{\bar f}{K}_{\beta\alpha_1...\alpha_I}
        + v^\gamma \partial_\gamma \ord{\bar f}{K}_{\beta\alpha_1...\alpha_I} )\right)v^{\alpha_1}...v^{\alpha_I}\\
        &=  \sum_{J=-2}^\infty  \frac{1}{c^{J}} \sum_{I=0}^{J+2} \frac{1}{I!}
         \partial_{\beta }\ord{\bar f}{J-I+2}_{\alpha_1...\alpha_I} v^{\alpha_1}...v^{\alpha_I}
        - \sum_{J=-1}^\infty \frac{1}{c^{J}} \sum_{I=0}^{J+1} \frac{1}{I!} (\partial_t \ord{\bar f}{J-I+1}_{\beta\alpha_1...\alpha_I}
        + v^\gamma \partial_\gamma \ord{\bar f}{J-I+1}_{\beta\alpha_1...\alpha_I} ) v^{\alpha_1}...v^{\alpha_I}
\end{split}\end{align}
Up to order $c^{-2}$ this expression reads
\begin{align}\begin{split}
        f_{\alpha\beta} \frac{d^2 x^\alpha}{d t^2}  &= c^2 \partial_\beta \ord{\Bf}{0} \\
        &+ c \left[ \partial_\beta \ord{\Bf}{1} - \partial_t \ord{\Bf}{0}_\beta + v^{\alpha_1}(\partial_\beta\ord{\Bf}{0}_{\alpha_1}  -  \partial_{\alpha_1} \ord{\Bf}{0}_\beta) \right]\\
        &+ c^0 \left[ \partial_\beta \ord{\Bf}{2} - \partial_t \ord{\Bf}{1}_\beta + v^{\alpha_1} (\partial_\beta \ord{\Bf}{1}_{\alpha_1} -  \partial_{\alpha_1} \ord{\Bf}{1}_\beta - \partial_t \ord{\Bf}{0}_{\beta\alpha_1} ) + v^{\alpha_1} v^{\alpha_2}  (\tfrac{1}{2}\partial_\beta \ord{\Bf}{0}_{\alpha_1\alpha_2} - \partial_{\alpha_1} \ord{\Bf}{0}_{\beta\alpha_2} )\right] \\
        &+ \frac{1}{c} \bigg[ \partial_\beta \ord{\Bf}{3} - \partial_t \ord{\Bf}{2}_\beta
        + v^{\alpha_1} (\partial_\beta \ord{\Bf}{2}_{\alpha_1} -  \partial_{\alpha_1} \ord{\Bf}{2}_\beta - \partial_t \ord{\Bf}{1}_{\beta\alpha_1} )
        + v^{\alpha_1} v^{\alpha_2}  (\tfrac{1}{2}\partial_\beta \ord{\Bf}{1}_{\alpha_1\alpha_2} -  \partial_{\alpha_1} \ord{\Bf}{1}_{\beta\alpha_2}  - \tfrac{1}{2}\partial_t \ord{\Bf}{0}_{\beta\alpha_1\alpha_2}) \\
        &+ v^{\alpha_1} v^{\alpha_2} v^{\alpha_3} (\tfrac{1}{3!}\partial_\beta \ord{\Bf}{0}_{\alpha_1\alpha_2\alpha_3} - \tfrac{1}{2}\partial_{\alpha_1} \ord{\Bf}{0}_{\beta\alpha_2\alpha_3} ) \bigg] \\
        &+ \frac{1}{c^2} \bigg[ \partial_\beta \ord{\Bf}{4} - \partial_t \ord{\Bf}{3}_\beta
        + v^{\alpha_1} (\partial_\beta \ord{\Bf}{3}_{\alpha_1} -  \partial_{\alpha_1} \ord{\Bf}{3}_\beta - \partial_t \ord{\Bf}{2}_{\beta\alpha_1} )
        + v^{\alpha_1} v^{\alpha_2}  (\tfrac{1}{2}\partial_\beta \ord{\Bf}{2}_{\alpha_1\alpha_2} -  \partial_{\alpha_1} \ord{\Bf}{2}_{\beta\alpha_2}  - \tfrac{1}{2}\partial_t \ord{\Bf}{1}_{\beta\alpha_1\alpha_2}) \\
        &+ v^{\alpha_1} v^{\alpha_2} v^{\alpha_3} (\tfrac{1}{3!}\partial_\beta \ord{\Bf}{1}_{\alpha_1\alpha_2\alpha_3} - \tfrac{1}{2}\partial_{\alpha_1} \ord{\Bf}{1}_{\beta\alpha_2\alpha_3} - \tfrac{1}{3!}\partial_t \ord{\Bf}{0}_{\beta\alpha_1\alpha_2\alpha_3})
        + v^{\alpha_1} v^{\alpha_2} v^{\alpha_3} v^{\alpha_4}(\tfrac{1}{4!}\partial_\beta \ord{\Bf}{0}_{\alpha_1\alpha_2\alpha_3\alpha_4} - \tfrac{1}{3!}\partial_{\alpha_1} \ord{\Bf}{0}_{\beta\alpha_2\alpha_3\alpha_4} )  \bigg] \\
        &+ \mathcal{O}(c^{-3})\,.
\end{split}\end{align}
Providing the inverse of the tensor $f_{\alpha\beta}$ order by order in $1/c,$ we can thus derive the covariant spatial acceleration equation.

Using the expansions given in Eqs. \eqref{eqn:LagExp} and \eqref{eqn:ExtFieldExp}, we find that the matrix $f_{\alpha\beta}$ can be expanded as
\begin{align}
	\begin{split}\label{eq:fddx=G}
    f_{\alpha\beta}
    &= \sum_{J=0}^\infty  \frac{1}{c^{J}} \sum_{I=0}^{J} \frac{1}{I!}\
    \ord{\bar f}{J-I}_{\alpha\beta\alpha_1...\alpha_I} v^{\alpha_1}...v^{\alpha_I},\\
    & = \ord{\bar f}{0}_{\alpha\beta}
    + \frac{1}{c} (\ord{\bar f}{1}_{\alpha\beta} + \ord{\bar f}{0}_{\alpha\beta\alpha_1} v^{\alpha_1})
    + \frac{1}{c^2} (\ord{\bar f}{2}_{\alpha\beta} + \ord{\bar f}{1}_{\alpha\beta\alpha_1} v^{\alpha_1} + \tfrac{1}{2} \ord{\bar f}{0}_{\alpha\beta\alpha_1\alpha_2} v^{\alpha_1}v^{\alpha_2} )\\
    &+ \frac{1}{c^3} (\ord{\bar f}{3}_{\alpha\beta} + \ord{\bar f}{2}_{\alpha\beta\alpha_1} v^{\alpha_1} + \tfrac{1}{2} \ord{\bar f}{1}_{\alpha\beta\alpha_1\alpha_2} v^{\alpha_1}v^{\alpha_2} + \tfrac{1}{3!} \ord{\bar f}{0}_{\alpha\beta\alpha_1\alpha_2\alpha_3} v^{\alpha_1}v^{\alpha_2} v^{\alpha_3}) + \mathcal{O}(c^{-4}),
    \end{split}
\end{align}
and can be inverted order by order. We use the ansatz
\begin{align}
    f^{\alpha\beta}
    &= \ord{\bar f}{0}^{\alpha\beta} + \sum_{J=1}^\infty \frac{1}{c^J} \ord{q}{J}^{\ \alpha\beta}\,
\end{align}
to determine the corrections $\ord{q}{J}^{\ \alpha\beta}$ to the desired order.

We write $d^2 x^\alpha / dt^2 = f^{\alpha\beta} G_{\alpha}$ schematically in orders of $c^{-1}$ as
\begin{align}\begin{split}\label{eq:acceq}
    \frac{d^2 x^\beta}{d t^2}
    &= \left(\ord{\bar f}{0}^{\alpha\beta} + \frac{1}{c} \ord{q}{1}^{\ \alpha\beta} + \frac{1}{c^2} \ord{q}{2}^{\ \alpha\beta} + \frac{1}{c^3} \ord{q}{3}^{\ \alpha\beta}+ \frac{1}{c^4} \ord{q}{4}^{\ \alpha\beta} + \mathcal{O}(c^{-5}) \right)\\
    &\times
    \left(c^2 \ord{G}{-2}_\alpha + c \ord{G}{-1}_\alpha + \ord{G}{0}_\alpha + \frac{1}{c} \ord{G}{1}_\alpha + \frac{1}{c^2} \ord{G}{2}_\alpha + \mathcal{O}(c^{-3}) \right),\\
    &= c^2 \ord{\bar f}{0}^{\alpha\beta}\ \ord{G}{-2}_\alpha
    + c \left(\ord{\bar f}{0}^{\alpha\beta}\ \ord{G}{-1}_\alpha + \ord{q}{1}^{\ \alpha\beta}\ \ord{G}{-2}_\alpha \right)\\
    &+ \left(\ord{\bar f}{0}^{\alpha\beta}\ \ord{G}{0}_\alpha + \ord{q}{1}^{\ \alpha\beta}\ \ord{G}{-1}_\alpha + \ord{q}{2}^{\ \alpha\beta}\ \ord{G}{-2}_\alpha\right)\\
    &+\frac{1}{c} \left(\ord{\bar f}{0}^{\alpha\beta}\ \ord{G}{1}_\alpha + \ord{q}{1}^{\ \alpha\beta}\ \ord{G}{0}_\alpha + \ord{q}{2}^{\ \alpha\beta}\ \ord{G}{-1}_\alpha + \ord{q}{3}^{\ \alpha\beta}\ \ord{G}{-2}_\alpha \right)\\
    &+ \frac{1}{c^2}\left(\ord{\bar f}{0}^{\alpha\beta}\ \ord{G}{2}_\alpha + \ord{q}{1}^{\alpha\beta}\ \ord{G}{1}_\alpha + \ord{q}{2}^{\alpha\beta}\ \ord{G}{0}_\alpha + \ord{q}{3}^{\alpha\beta}\ \ord{G}{-1}_\alpha + \ord{q}{4}^{\alpha\beta}\ \ord{G}{-2}_\alpha \right)\\
    &+ \mathcal{O}(c^{-3})\,,
\end{split}\end{align}
in order to identify the dominant terms contributing to the nonrelativistic limit. Lower-order terms in $f^{\alpha\beta}$ only contribute to lower-order terms in the acceleration. The terms $\ord{G}{J}_\alpha$ can be read of from the expansion \eqref{eq:fddx=G}. The important assumption in this expansion is that $f^{\alpha\beta}$ contains no terms proportional to positive powers of $c$.

We want to ensure a well-defined $c\to\infty$ limit in the nonrelativistic acceleration equations, which imposes two constraints
\begin{align}\label{eq:accneccond}
    \ord{G}{-2}_\alpha = 0\,,\quad \ord{G}{-1}_\alpha = 0\,.
\end{align}
These constraints can be solved by setting
\begin{align}
     \partial_\beta \ord{\Bf}{0} = 0\,,
    \quad
    \ord{\Bf}{0}_\alpha = \partial_\alpha \Psi \,,
    \quad
    \ord{\Bf}{1} = \partial_t \Psi + a \,,
\end{align}
where $\Psi = \Psi(t,x)$ is a potential function and $a$ a constant.

Applying the constraints in \eqref{eq:accneccond}, we can simplify the acceleration equation, \eqref{eq:acceq}, to order $c^{-2}$ such that
\begin{align}\label{eq:accnonrelsimp}
\boxed{
    \frac{d^2 x^\beta}{d t^2}
    = \ord{\bar f}{0}^{\alpha\beta}\ \ord{G}{0}_\alpha
    +\frac{1}{c} \left(\ord{\bar f}{0}^{\alpha\beta}\ \ord{G}{1}_\alpha + \ord{q}{1}^{\ \alpha\beta}\ \ord{G}{0}_\alpha \right)
    + \frac{1}{c^2}\left(\ord{\bar f}{0}^{\alpha\beta}\ \ord{G}{2}_\alpha + \ord{q}{1}^{\ \alpha\beta}\ \ord{G}{1}_\alpha + \ord{q}{2}^{\ \alpha\beta}\ \ord{G}{0}_\alpha  \right)
    + \mathcal{O}(c^{-3})\,.}
\end{align}
Thus only $\ord{q}{1}^{\ \alpha\beta}$ and $\ord{q}{2}^{\ \alpha\beta}$ are required to this order and given by
\begin{align}
    \ord{q}{1}^{\ \sigma\gamma}
    &= - \ord{\bar f}{0}^{\sigma\beta} \ord{\bar f}{0}^{\alpha\gamma} (\ord{\bar f}{1}_{\alpha\beta} + \ord{\bar f}{0}_{\alpha\beta\alpha_1} v^{\alpha_1})\,,\\
    \ord{q}{2}^{\ \sigma\gamma}
    &=  \ord{\bar f}{0}_{\alpha\lambda}\ \ord{q}{1}^{\ \alpha\gamma}\ \ord{q}{1}^{\ \sigma\lambda}
    - \ord{\bar f}{0}^{\sigma\beta} \ord{\bar f}{0}^{\alpha\gamma} (\ord{\bar f}{2}_{\alpha\beta} + \ord{\bar f}{1}_{\alpha\beta\alpha_1} v^{\alpha_1} + \tfrac{1}{2}\ord{\bar f}{0}_{\alpha\beta\alpha_1\alpha_2} v^{\alpha_1}v^{\alpha_2})\,.
\end{align}
In the nonrelativistic limit, $c\to\infty$, the spatial acceleration equation, \eqref{eq:accnonrelsimp}, reads explicitly
\begin{align}\label{eq:accNewt}
        &\ord{\bar f}{0}_{\alpha\beta} \ord{\frac{d^2 x^\alpha}{d t^2}}{0}
        =  \ord{G}{0}_\beta
        =
        \partial_{\beta }\ord{\bar f}{2} - \partial_t \ord{\bar f}{1}_{\beta }
        + v^{\sigma_1 } \left[\partial_{\beta }\ord{\bar f}{1}_{\sigma_1} - \partial_{\sigma_1}\ord{\bar f}{1}_{\beta} - \partial_t \ord{\bar f}{0}_{\sigma_1\beta} \right]
        + v^{\sigma_1 }v^{\sigma_2 } \left[\tfrac{1}{2} \partial_{\beta }\ord{\bar f}{0}_{\sigma_1\sigma_2} -  \partial_{\sigma_1 }\ord{\bar f}{0}_{\beta\sigma_2}\right]\,.
\end{align}
The orders $c^{-1}$ and $c^{-2}$ are lengthy expression which we display explicitly in appendix \ref{app:acceqExpPPN}.

Since we are studying the nonrelativistic slow-velocity limit of general Finslerian particle motion, we have not imposed the conventional  (p)PN constraints for a metric, \eqref{eq:normalPN}, which lead to the vanishing of the order-$c^{-1}$ correction in the Lagrangian \eqref{eqn:PertLag}.

For the $c^0$ order we observe that several contributions appear. Their interpretation depends on the physical system under consideration and, in the end, on the field equations which determine the point-particle action \eqref{eq:ppact} or its constituents:
\begin{itemize}
    \item The scalar potential $\ord{\bar f}{2}$, which may contain the Newtonian gravitational potential or the electromagnetic scalar potential.
    \item The vector potential $\ord{\bar f}{1}_{\alpha}$, which may contain the vector electromagnetic vector potential or non-classical Newtonian gravitational field contributions.
    \item The derivatives of the 2-tensor modes (in the Riemannian case the background metric) $\partial_\beta \ord{\bar f}{0}_{\alpha\gamma}$, which classically  encode fictitious forces.
\end{itemize}
Whether, or not, some of these contributions are set to certain values \textit{a priori}, such as $\ord{\bar f}{1}_{\alpha} = 0 = \partial_\beta \ord{\bar f}{0}_{\alpha\gamma}$, as in the usual (p)PN-analysis, depends on the specific questions being investigated.

Next, we will use the nonrelativistic expansion to investigate the WEP for MDRs.

\section{Modified dispersion relations and the weak equivalence principle}\label{sec:EP}
The spatial acceleration equation \eqref{eq:acceqallorders} is the perfect starting point to investigate whether, or not, MDRs are compatible with the WEP. The WEP is best known as the equality of gravitational and inertial masses (see \cite{Ohanian:1977hap}), such that free fall of test bodies is independent of their mass $M$. As this formulation heavily draws on Newtonian concepts, we will concentrate on the, more careful formulation quoted in the introduction \cite{Will:2018bme}.

For the specific context of the present work, the WEP can only be valid if the spatial acceleration equation \eqref{eq:acceqallorders} is independent of the particle's mass. We will see in this section that for generic MDRs this is not the case, and we will identify criteria MDRs have to satisfy such that the WEP is valid.

For this purpose, we first recall how MDRs are described as Hamilton functions and how the motion of particles subject to an MDR can be described by Finsler geodesics. This will allow us to formulate Theorem \ref{thm:WEPMDR} which summarises the necessary and sufficient condition for an MDR to satisfy the WEP. Afterwards we will quantify the deviation from the WEP by calculating the E\"otv\'os factor for polynomial MDRs in section \ref{eq:eot} explicitly, and in particular for the $\kappa$-Poincar\'e dispersion relation  in the bicrossproduct basis as example. Finally we identify modified dispersion relation which lead to particle motion satsifying the WEP in Section \ref{ssec:MDRsWEP}.

\subsection{The spatial acceleration equation}\label{ssec:MDRacc}
In order to investigate the WEP for MDRs we consider the Finsler function that encodes the point-particle motion of free particles subject to a dispersion relation of the form
\begin{align}\label{eq:MDR}
    H(x,p) = g^{ab}(x)p_ap_b +\Xi h(x,p) = M^2\,.
\end{align}
Here $g$ denotes a Lorentzian spacetime metric, $p$ denote momenta of particles, $h$ is a perturbation function on the $1$-particle phase space and $\Xi$ the perturbation parameter. In the context of quantum gravity phenomenology $\Xi$ is often considered to be a power of the inverse Planck energy~\cite{Addazi:2021xuf}.

Employing the Helmholtz action of free point particles, we obtain the corresponding Finsler function to first order in $\Xi$, see \cite{Lobo:2020qoa} for the derivation,
\begin{align}
    F = \sqrt{g_{ab}(x)\dot x^a \dot x^b} \left( 1 - \frac{\Xi}{2 M^2} h(x,\bar p)\right)\,,\quad
    \bar p_c = M \frac{g_{cd}(x)\dot x^d}{\sqrt{g_{ab}(x)\dot x^a \dot x^b}}\,.
\end{align}
In the language of this paper, this leads to a spatial Lagrangian $f$, defined in \eqref{eq:ppact}, of the form
\begin{align}\label{eq:fMDR}
    f(ct, x^\alpha, \epsilon^\alpha) = \mathfrak{f} \left( 1 - \frac{\Xi}{2 M^2} \bar h(ct, x^\alpha,\epsilon^\alpha )\right)\,,
\end{align}
where we use the abbreviations
\begin{align}
    \mathfrak{f} := \sqrt{g_{00} + 2 g_{0\alpha}\epsilon^\alpha + g_{\alpha\beta}\epsilon^\alpha\epsilon^\beta}\,,\quad
    \bar h(ct, x^\alpha,\epsilon^\alpha ) := h(ct, x^\alpha,\bar p(ct,x^\alpha,\epsilon^\alpha))\,,\quad
    \bar p_c :=  \frac{ M(g_{c0} + g_{c\alpha}\epsilon^\alpha)}{\mathfrak{f}}\,.
\end{align}
Thus, we obtain the spatial acceleration \eqref{eq:acceqallorders} to first order in $\Xi$
\begin{align}\label{eq:accMDR}
    \mathfrak{f}{}_{\beta\alpha} \frac{d^2 x^\beta}{dt^2}
    = a^{\rm GR}_\alpha
    - \frac{\Xi}{2M^2}\left[\bar h\ a^{\rm GR}_\alpha -  \mathfrak{f}^{\sigma\gamma} (\mathfrak{f} \bar h)_{\alpha\sigma} a^{\rm GR}_\gamma + c^2 \mathfrak{f} \partial_\alpha\bar h -c (\mathfrak{f}_\alpha (\partial_t \bar h+v^\gamma\partial_\gamma \bar h) + \partial_t (\mathfrak{f} \bar h_\alpha) + v^\gamma\partial_\gamma (\mathfrak{f} \bar h_\alpha)) \right]\,,
\end{align}
where we used the first-order inverse of $f_{\alpha\beta}$
\begin{align}
        f^{\alpha\beta} = \mathfrak{f}^{\alpha\beta} + \frac{\Xi}{2M^2}\mathfrak{f}^{\alpha\sigma} \mathfrak{f}^{\beta\rho} (\mathfrak{f} \bar h)_{\sigma\rho}\,,
    \end{align}
introduced the generally relativistic acceleration
\begin{equation}
    a^{\rm GR}_\alpha=c^2 \partial_\alpha \mathfrak{f} - c ( \partial_t \mathfrak{f}_{\alpha} +v^\gamma  \partial_\gamma \mathfrak{f}_{\alpha} )\,,
\end{equation}
and remark that further derivatives acting on $\bar h$ can be expanded according to its definition in terms of the original perturbation function $h(x,p)$ as
\begin{align}
    \partial_\alpha \bar h = \partial_\alpha h+\bar\partial^c h\ \partial_\alpha\bar p_c\,,\quad
    \partial_t \bar h = \partial_t h+\bar\partial^c h\ \partial_t\bar p_c\,.
\end{align}

Observe that $\mathfrak{f}$ does not depend on $M$. The perturbation function $\bar h=h(x,\bar p)$ depends on $M$ through $\bar p$, where $\bar p = M \hat p$ and $\hat p$ is independent of $M$. Hence, in general the spatial acceleration of a test body depends on the particle's mass, which signals a violation of the WEP. In order to implement the WEP the factors of $M$ must cancel from the spatial acceleration equation \eqref{eq:accMDR}. This can be ensured in two ways.

Either, the product $\Xi h$ has a specific scaling behaviour w.r.t the mass or the term in square brackets in \eqref{eq:accMDR} vanishes. We call an MDR for which the latter case holds classically ``dynamically trivial'', since it does not influence the point-particle motion at all. Assume a dynamically non-trivial MDR and consider a perturbation $h$ that is $n$-homogeneous in $\bar p$, meaning that $h(x,\lambda \bar p) = \lambda^n h(x,\bar p)$. This implies that $\bar h=h(x, \bar p) = M^n h(x, \hat p)$ and $h(x, \hat p)$ is independent of $M$. The scaling behaviour of the derivatives of $\bar h$ follow immediately. This  homogeneity assumption is justified, since most MDRs studied in the literature lead, when studied as perturbation of GR, to monomials of some degree $n$ in $p$ \cite{Jacob:2008bw,Amelino-Camelia:2008aez,Addazi:2021xuf,Laanemets:2022rmn,Amelino-Camelia:2023srg}. Then, in order to eliminate the mass dependence from the acceleration, one needs to set $\Xi = \Xi(M) = M ^{2-n} \hat \Xi$, where $\hat \Xi$ is independent of $M$.

Let us summarize this important result:
\begin{thm}[Modified dispersion relations and the weak equivalence principle]\label{thm:WEPMDR}
    Let $H$ be a Hamilton function on the 1-particle phase space of a curved spacetime $(M,g)$ which defines classically dynamically non-trivial modified dispersion relation
    \begin{align}
        H(x,p) = g^{ab}(x)p_ap_b + \Xi(M) h(x,p) = M^2\,.
    \end{align}
    Then the weak equivalence principle in the form,
    \begin{quote}
        "The trajectory of a freely falling “test” body (one not acted upon by such forces as electromagnetism and too small to be affected by tidal gravitational forces) is independent of its internal structure and composition",
    \end{quote}
    is valid if and only if the product of the perturbation function with the perturbation coupling parameter $\Xi h$ is homogeneous of degree $2$ with respect to the mass $M$. For an $n$-homogeneous perturbation function
    \begin{align}\label{eq:2homh}
        h(x,\lambda p) = \lambda^n h(x,p)\,,
    \end{align}
    this implies that the coupling constant is given by $\Xi = M^{2-n} \hat \Xi$, where $\hat \Xi$ is a dimensionless number and independent of $M$.
\end{thm}
As a consequence of this theorem, the WEP can only be valid for mass independent coupling $\Xi(M) = \hat \Xi$ if the perturbation function  is homogeneous of degree $n=2$ with respect to the momenta $p$. In any other case, either the WEP is violated or $\Xi$ must depend on the particle's mass.

The necessity of a mass dependent coupling between different terms in the Hamilton function, which determines the dynamics of point particles to ensure the WEP, is less unusual than it might sound at first glance. Already in classical non-relativistic Newtonian gravity, this is the case. Let $\phi$ be the Newtonian gravitational potential, and $\vec p$ the momentum of a non-relativistic classical point particle. Then the Hamiltonian which describes the motion of the particle in the gravitational field is given by
\begin{align}
	H(\vec x, \vec p) = \frac{\vec p^2}{2M} + M \phi \,,
\end{align}
and the WEP is valid. If the factors $1/2M$ and $M$ in front of the $\vec p^2$ term and the term independent of $\vec p$ did not depend on $M$ in the way they do, the WEP would be violated. Thus, even in non-relativistic Newtonian gravity, it is essential that the coupling between the Newtonian potential and the point particle is not universal for all particles, but depends on their mass in a specific way. Hence, this option should be taken into account when studying MDRs in the context of quantum gravity.

In particular, in deformed relativity (DSR) models, a mass dependence of $\Xi$, has to emerge for macroscopic objects in some or the other to avoid the soccer ball problem \cite{Amelino-Camelia:2002uql,Hossenfelder:2007fy,Hossenfelder:2014ifa,Amelino-Camelia:2014gga}, even if it is not there at the level of the fundamental constituents. In SI-units the Planck mass amounts to $\sim 10^{-5}$g, such that quantum-gravity effects should be commonplace if $\Xi$ did not scale with an inverse power of the number of constituents $N\sim M/M_{\rm c},$ with a mass scale associated with the elementary constituents $M_{\rm c}$. Further details of this argument are discussed in Appendix \ref{app:CompPart} and will be needed when we compare the $\kappa$-Poincar\'e dispersion relation with the microscope measurements in equation \eqref{eq:micbounds}.

To clarify the implications of these observations we discuss next how to obtain the E\"otv\'os factor for a specific type of MDR, which leads to a connection between constraints on MDRs that do not satisfy \eqref{eq:2homh} and tests of the WEP.

\subsection{E\"otv\'os factor for MDRs}\label{eq:eot}
Violations of the WEP are conventionally quantified by virtue of the E\"otv\'os factor $\eta$, which measures the difference between the gravitational acceleration of two distinct test bodies in one and the same gravitational field and motional state
\begin{align}
    \eta = 2 \frac{|\vec a_1 - \vec a_2|}{|\vec a_1 + \vec a_2|}\,.\label{eq:DefEotvos}
\end{align}
For motion determined by the generally relativistic spatial acceleration equation this factor vanishes and hence the WEP is satisfied. Thus, for MDRs of the type \eqref{eq:MDR}, the absolute values appearing in \eqref{eq:DefEotvos} can approximately be taken with respect to the spatial metric since possible corrections will necessarily be of next-to-leading order (order $\Xi^2$ in the language of the preceding section). Further, recalling that the mass \(M\) of the pointlike test body is the sole property which determines the acceleration~\eqref{eq:accMDR}, we will consider two test bodies of distinct mass \(M_1\) and \(M_2\).

Consider a perturbation to the general-relativistic acceleration such that $\vec{a}=\vec{a}_{\rm GR}+ \Delta \vec{a}.$ Then to first order in the perturbation $\Delta \vec{a},$ the E\"otv\'os factor assumes the form
\begin{equation}
    \eta\simeq \frac{|\Delta \vec{a}(M_1)-\Delta \vec{a}(M_2)|}{|\vec{a}_{\rm GR}|}.
\end{equation}
To determine this observable quantitatively, we assume again that the perturbation function $h(x,p)$ is homogeneous of degree $n$ in $p$. We already found in Theorem \ref{thm:WEPMDR}, that if $n\neq 2$ the underlying dispersion relation violates the WEP, if the mass dependence is not counterbalanced by the coupling constant. To trace this violation we can introduce a function $\hat{h}$ which is independent of $M$ via
\begin{equation}
    h(x,\bar p) = M^n h(x,\hat p) = M^n \hat h\,,
    \quad
    \hat p_c = \frac{g_{cd}(x)\dot x^d}{\sqrt{g_{ab}(x)\dot x^a \dot x^b}}\,.
\end{equation}
From \eqref{eq:accMDR} we conclude that we can similarly express the perturbation of the acceleration as
\begin{equation}
    \Delta\vec{a}=\Xi(M) M^{n-2}\Delta\vec{\hat{a}},
\end{equation}
where $\Delta\vec{\hat{a}}$ is independent of $M.$ Then, the E\"otv\'os factor reads
\begin{equation}
    \eta=  \left| \Xi(M_1)M_1^{n-2}-\Xi(M_2)M_2^{n-2}\right|  \frac{|\Delta\vec{\hat{a}}|}{|\vec{a}_{\rm GR}|}.
\end{equation}

To provide an explicit example, we further derive the factor $|\Delta\vec{\hat{a}}|$ for a perturbation function $\hat h$ of the type
\begin{align}
    \hat h = h^{a_1...a_n}(x)\hat p_{a_1} ... \hat p_{a_n}\,.
\end{align}
To demosntrate the appearence of the violation of the WEP, we derive the nonrelativistic acceleration in the Newtonian regime and at first order in $\Xi=\Xi(M)$. The derivation can straightforwardly be generalized to any pN order.

We first observe that for MDRs the function $f$ is given by
\begin{align}\label{eq:fMDR1}
    f = \mathfrak{f}\left(1 - \frac{\Xi}{2 M^2} M^n h^{a_1...a_n} \frac{( g_{a_10} + g_{a_1\alpha_1} \epsilon^{\alpha_1}){}_{...}( g_{a_n0} + g_{a_n\alpha_n} \epsilon^{\alpha_n}) }{\mathfrak{f}^{n}}\right)\,.
\end{align}
Assuming that for $\Xi=0$, the usual Newtonian limit is valid, the background metric satisfies \eqref{eq:normalPN}. Thus, we can approximately express the mass-independent part of the correction $\hat{h}$ as
\begin{equation}
    \hat{h} \equiv h^{a_1...a_n} \frac{((1+2\frac{\phi}{c^2}) \delta_{a_10} + \delta_{a_1\alpha_1} \epsilon^{\alpha_1}){}_{...}( (1+2\frac{\phi}{c^2}) \delta_{a_n0} + \delta_{a_n\alpha_n} \epsilon^{\alpha_n}) }{\mathfrak{f}^{n}}\,.
\end{equation}
The relevant contributions to the nonrelativistic limit of the acceleration equation, \eqref{eq:acceq}, read
\begin{align}
    \ord{\Bf}{2} &= \phi - \Xi \frac{M^{n-2}}{2} \left[2 n \phi\ \ord{h}{0}^{0...0}  + \ord{h}{2}^{0...0} \right]
    \,,\\
    \ord{\Bf}{1}_\alpha &=  \Xi \frac{M^{n-2}}{2} (n-1)^2\ \ord{h}{1}^{\beta 0...0} \delta_{\beta\alpha}\,,\\
    \ord{\Bf}{0}_{\alpha\beta} &= - \delta_{\alpha\beta} - \Xi \frac{M^{n-2}}{2}  \left[ 2 \binom{n}{2} \ord{h}{0}^{\sigma_1\sigma_20...0} \delta_{\sigma_1\alpha} \delta_{\sigma_2\beta} - 2 (n-1) \ord{h}{0}^{0...0} \delta_{\alpha\beta} \right]\,.
\end{align}
Besides the Newtonian potential $\phi$, polynomial MDRs of monomial type generically introduce four new potentials at the Newtonian level $\ord{h}{0}^{0...0}, \ord{h}{0}^{\alpha\beta...0}, \ord{h}{1}^{\alpha...0}, \ord{h}{2}^{0...0}$, which in general may depend on all coordinates $(t,x^\alpha)$.

In order to examine the E\"otv\"os factor, we introduce the short-hand notation
\begin{align}
    \ord{\Bf}{2} &= \phi - \Xi \frac{M^{n-2}}{2} \ord{Q}{2}\,,\\
    \ord{\Bf}{1}_\alpha &=  \Xi \frac{M^{n-2}}{2} \ord{Q}{1}_\alpha\,,\\
    \ord{\Bf}{0}_{\alpha\beta} &= - \delta_{\alpha\beta} - \Xi \frac{M^{n-2}}{2}  \ord{Q}{0}_{\alpha\beta}\,,
\end{align}
which leads to a correction of the Newtonian acceleration of the form
\begin{align}
    \Delta\vec{\hat{a}}^\alpha = \Xi \frac{M^{n-2}}{2} K^\alpha\,,
\end{align}
with
\begin{align}\label{eq:DefK}
    K^\alpha
    = \partial^\alpha \ord{Q}{2} + \partial_t\ord{Q}{1}^\alpha + \ord{Q}{0}^{\alpha\beta}\partial_\beta \phi
    - v^{\sigma}(\partial_\beta \ord{Q}{1}_\sigma - \partial_\sigma \ord{Q}{1}_\beta + \partial_t\ord{Q}{0}_{\sigma\beta})\delta^{\alpha\beta}
    + v^{\sigma_1} v^{\sigma_2}(\tfrac{1}{2}\partial_\beta \ord{Q}{0}_{\sigma_1\sigma_2} - \partial_{\sigma_1} \ord{Q}{0}_{\beta\sigma_2})\delta^{\alpha\beta}\,.
\end{align}
To first order in $\Xi$, the E\"otv\'os factor is now easily expressed as
\begin{align}\label{eq:eotMDR}
    \eta = |M_1^{n-2} \Xi(M_1) - M_2^{n-2} \Xi(M_2)| \frac{1}{2} \frac{\sqrt{\delta_{\alpha\beta}K^\alpha K^\beta} }{\sqrt{\delta^{\alpha\beta}\partial_\alpha \phi \partial_\beta \phi}}\,,
\end{align}
where we allowed that $\Xi = \Xi(M)$ may depend on the particle's mass. We see explicitly that polynomial MDRs lead to a deviation from the WEP already at Newtonian order, or $\Xi$ must be of the form $\Xi = M^{2-n} \hat{\Xi}$, as we found out in Theorem \ref{thm:WEPMDR}, or $K^\alpha$ must vanish. The last point would mean that the MDR is dynamically trivial.

Thus, measurements of the E\"otv\'os factor between different masses immediately impose bounds on the dispersion relation at nonrelativistic Newtonian order or imply a specific scaling behaviour of the modification term $\Xi h$ in the Hamiltonian with respect to the mass $M$. By analogy with our presentation at the Newtonian level, the deviation can be quantified and constraint at any pN order. To perform this derivation explicitly is ongoing work and opens a window to test MDRs by means of WEP tests at any desired pN-order.

After the general discussion, we apply our findings to the $\kappa$-Poincar\'e dispersion relation in the bicrossproduct basis on curved spacetime \cite{Barcaroli:2017gvg}. As the rest of our paper, the following analysis relies on the assumption that the dispersion relation fully determines the motion of point particles via the Hamilton equations of motion. Since we are considering just single particles, the multi-particle sector of the full $\kappa$-Poincar\'e framework \cite{Lukierski:1992dt,Majid:1994cy,Kowalski-Glikman:2002iba,Amelino-Camelia:2000stu}, deformed Lorentz transformations, or modified energy-momentum addition laws, which are essential to DSR theories, are not taken into account here. Their influence on the WEP goes beyond the scope of the present investigation.

The $\kappa$-Poincar\'e dispersion relation we are considering is defined by a spacetime metric $g$ and a unit timlike vector field $Z$ (meaning in our convention that $g(Z,Z) = g^{ab}Z_aZ_b=1$). The lowest order corrections to GR are given by
\begin{align}\label{eq:MDRkappa}
    h^{abc}(ct, x^\alpha) = \frac{1}{3} (g^{ab}Z^c + g^{cb}Z^a + g^{ac}Z^b) - Z^a Z^b Z^c\,.
\end{align}
Using this ansatz to construct the spatial point-particle Lagrangian \eqref{eq:fMDR1}, we obtain (in this equation indices are raised and lowered with the spacetime metric $g$)
\begin{align}
	f
	&= \mathfrak{f}  - \Xi \tfrac{M}{2} \bigg[  \left(g_{00}Z_0 - Z_0^3\right) + \epsilon^\alpha \left(2 g_{\alpha0}Z_0 + g_{00}Z_\alpha -  3 Z_\alpha Z_0^2\right)\,,\\ &\quad\qquad\qquad+\epsilon^\alpha \epsilon^\beta \left(g_{\alpha\beta}Z_0 + 2 g_{\alpha0}Z_\beta -  3 Z_\alpha Z_\beta Z_0 \right)+ \epsilon^\alpha \epsilon^\beta \epsilon^\gamma \left(g_{\alpha\beta}Z_\gamma - Z_\alpha Z_\beta Z_\gamma\right) \bigg] \,.
\end{align}
This expression can be expanded to Newtonian order ($c^0$) by using the usual pN parametrization of the metric \eqref{eq:normalPN} and imposing that $\ord{Z}{0}^{\ 0} = 1$. Solving the normalisation condition $g_{ab}Z^{a}Z^b=1$ order by order, up to order $1/c^2$, this amounts to the constraints $\ord{Z}{0}^\alpha=0=\ord{Z}{1}^0$, and $\ord{Z}{2}^0=- \frac{1}{2}\ord{g}{2}{}_{00} + \delta_{\alpha\beta}\ord{Z}{1}^\alpha\ord{Z}{1}^\beta/2$. As a result, we find
\begin{align}\label{eq:fkappa}
	f =   c^2 +  \left( \tfrac{1}{2} \ord{g}{2}{}_{00} - \tfrac{1}{2} v^2  \right) - \frac{\Xi M}{2}\left(2 \ord{g}{2}{}_{00}  - \delta_{\alpha\beta}\ord{Z}{1}^\alpha\ord{Z}{1}^\beta  -2\ord{Z}{1}_\alpha v^\alpha -  v^2\right)\,,
\end{align}
from which one obtains
\begin{align}\label{eq:kappa2}
	    \ord{\Bf}{2} &= \phi - \Xi \frac{M}{2} \left(4 \phi - \delta_{\alpha\beta} \ord{Z}{1}^\alpha\ord{Z}{1}^\beta \right)\,, &\Rightarrow&\qquad \ord{Q}{2} = 4\phi - \delta_{\alpha\beta} \ord{Z}{1}^\alpha\ord{Z}{1}^\beta \,,\\
		\ord{\Bf}{1}_\alpha &=\Xi M\ord{Z}{1}_\alpha \,,&\Rightarrow&\qquad \ord{Q}{1}_\alpha= 2\ord{Z}{1}_\alpha\,,\\
		\ord{\Bf}{0}_{\alpha\beta} &= - \delta_{\alpha\beta} + \Xi M \delta_{\alpha\beta} ,&\Rightarrow&\qquad \ord{Q}{0}_{\alpha\beta} = -2  \delta_{\alpha\beta}\,.
\end{align}
Alternatively, we can also expand the perturbation \eqref{eq:MDRkappa} directly into the pN orders as
\begin{align}
	\ord{h}{0}^{000} &
	= \ord{Z}{0}^{\ 0}(1 - (\ord{Z}{0}^{\ 0})^2)
	=0 \,,\\
	\ord{h}{2}^{000} &
	= 2 \phi \ord{Z}{0}^{\ 0} + \ord{Z}{2}^{\ 0}(1 -3  (\ord{Z}{0}^{\ 0})^2 )- 3 \ord{Z}{0}^{\ 0} (\ord{Z}{1}^{\ 0})^2
	=  4\phi - \delta_{\alpha\beta} \ord{Z}{1}^\alpha\ord{Z}{1}^\beta\,,\\
	\ord{h}{1}^{\alpha00} &
	= \ord{Z}{1}^{\ \alpha} \left( \tfrac{1}{3}  - (\ord{Z}{0}^{\ 0})^2\right) - 2\ord{Z}{0}^{\ \alpha}  \ord{Z}{0}^{\ 0} (\ord{Z}{1}^{\ 0})
	= - \frac{2}{3} \ord{Z}{1}^{\ \alpha} \,,\\
	\ord{h}{0}^{\alpha\beta0}&
	= - \tfrac{1}{3} \delta^{\alpha\beta}\ord{Z}{0}^{\ 0} - \ord{Z}{0}^{\ \alpha} \ord{Z}{0}^{\ \beta} \ord{Z}{0}^{\ 0}
	= - \frac{1}{3} \delta^{\alpha\beta}\,.
\end{align}
The additional terms in the spatial acceleration are then sourced by
\begin{align}
	\begin{split}
	K^\alpha
	& = - \partial^\alpha \left(  - 2 \phi + \delta_{\alpha\beta}\ord{Z}{1}^\alpha\ord{Z}{1}^\beta\right) - 2 v^\sigma \left( \partial^\alpha \ord{Z}{1}_\sigma -  \partial_\sigma \ord{Z}{1}^\alpha  \right) + 2 \partial_t  \ord{Z}{1}^\alpha\,,\\
	 &= - \partial^\alpha \left( - 2 \phi + \delta_{\alpha\beta}\ord{Z}{1}^\alpha\ord{Z}{1}^\beta+ 2 v^\sigma \ord{Z}{1}_\sigma\right)  + 2 \frac{d}{dt} \ord{Z}{1}^\alpha \,,\\
	 &= \delta^{\alpha\beta} \left( \frac{d}{dt} \frac{\partial}{\partial v^\beta} \left( - 2 \phi + \delta_{\alpha\beta}\ord{Z}{1}^\alpha\ord{Z}{1}^\beta + 2 v^\sigma \ord{Z}{1}_\sigma\right)- \frac{\partial}{\partial x^\beta} \left(  - 2 \phi + \delta_{\alpha\beta}\ord{Z}{1}^\alpha\ord{Z}{1}^\beta+ 2 v^\sigma \ord{Z}{1}_\sigma\right)     \right)\,,\\
	 &\equiv\delta^{\alpha\beta} \left( \frac{d}{dt} \frac{\partial\mathcal{Q}}{\partial v^\beta}- \frac{\partial\mathcal{Q}}{\partial x^\beta} \right) \,.
	 \end{split}
\end{align}
Thus, $K^\alpha$ is, not surprisingly according to equation \eqref{eq:geodspat}, the Euler-Lagrange expression derived from some function $\mathcal{Q}.$ Hence, $K^\alpha = 0$ implies, by the theory of Lagrangians and Euler-Lagrange equations \cite{Krupka:2015:gvg}, that the Lagrangian $\mathcal{Q}$ is a total derivative.

In summary, we just demonstrated the consequences of Theorem \eqref{thm:WEPMDR} on the example of a $\kappa$-Poincar\'e modification of general relativity. The corresponding Lagrangian \eqref{eq:kappa2} reads
\begin{align}
	f =   c^2 +  \tfrac{1}{2}\left(  \ord{g}{2}{}_{00} -  v^2  \right) \left(1 - \Xi M \right) + \tfrac{\Xi M}{2} \mathcal{Q}
	= c^2 + D   \left(  \ord{g}{2}{}_{00} -  v^2  \right)  + \tfrac{\Xi M}{2} \mathcal{Q}\,,\label{eqn:NRLagkappaPoincare}
\end{align}
where $\mathcal{Q} = - 2 \phi + \delta_{\alpha\beta}\ord{Z}{1}^\alpha\ord{Z}{1}^\beta + 2 v^\sigma \ord{Z}{1}_\sigma$ and $D$ is some constant. It either violates the WEP, or requires a mass depended coupling $\Xi=\hat{\Xi}/M$, or $K^\alpha=0$. In the latter case the model is classically dynamically trivial, since then \(\mathcal{Q}\) is a total derivative, \(\mathcal{Q} = d\mathfrak{Q}/dt\). The dynamics are equivalent to Newtonian gravity satisfying the WEP. The (possible) violation of the WEP is quantified by the E\"otv\'os factor
\begin{equation}\label{eq:eotkappa}
	\eta=\frac{1}{2}|\Xi(M_1) M_1- \Xi(M_2)M_2|\frac{|\vec{\nabla}( \ord{\vec{Z}}{1} \cdot \ord{\vec{Z}}{1} - 2 \phi )+2\vec{v}\times(\vec{\nabla}\times \ord{\vec{Z}}{1})-2\partial_t\ord{\vec{Z}}{1}|}{|\vec{\nabla}\phi|}\,,
\end{equation}
in classical vector space calculus notation.

We can use Eq. \eqref{eq:eotkappa} to constrain $\Xi,$ the model parameter characterizing the $\kappa$-Poincar\'e dispersion relation in the bicrossproduct representation, under certain, physically plausible, assumptions. The numerator of Eq. \eqref{eq:eotkappa} contains a term $\vec{\nabla}\phi$ and contributions depending on $\ord{Z}{1}^\alpha$. While $\vec{\nabla}\phi$ is dependent on the mass of the Earth and the distance from its centre, $\ord{Z}{1}^\alpha$ measures the velocity of the Lorentz deforming vector, a purported property of the universe, relative to the worldline along which we have taken the nonrelativistic limit. Generically, both of these quantities are independent of each other, and thus it is implausible that they are of the same order of magnitude for all different kinds of measurements of the WEP that exist -- specifically table-top and space-based experiments. In other words, either $|\ord{\vec{Z}}{1}|\ll|\vec{\nabla}\phi|$ or $|\ord{\vec{Z}}{1}|\gg|\vec{\nabla}\phi|.$ Therefore, it is reasonable to expect that
 \begin{equation}
 	|\vec{\nabla}( \ord{\vec{Z}}{1} \cdot \ord{\vec{Z}}{1} - 2 \phi )+2\vec{v}\times(\vec{\nabla}\times \ord{\vec{Z}}{1})-2\partial_t\ord{\vec{Z}}{1}|\geq 2|\vec{\nabla}\phi|.
 \end{equation}
Thus, the E\"otv\'os parameter satisfies
\begin{equation}
	\eta\geq|\Xi(M_1)M_1-\Xi(M_2)M_2|.\label{eq:EotvosGen}
\end{equation}
Hence, either 
\begin{equation}
	\Xi(M) = \hat \Xi / M\label{eq:WEPSavingScaling}
\end{equation} for some dimensionless parameter $\hat \Xi$ or the WEP is violated. At the level of elementary particles in quantum gravity phenomenology the parameter $\Xi=\hat \Xi$ is typically chosen to be independent of the particle's mass and the E\"otv\'os parameter becomes
\begin{equation}\label{eq:eotkappavanxi}
	\eta\geq \hat\Xi|M_1-M_2|\,.
\end{equation}

To save the WEP for corrections of third order in the four-momentum the mass dependence of the coupling has to be of the kind \eqref{eq:WEPSavingScaling}. Such a correction is rather unconventional in the context of quantum gravity phenomenology because it does not obviously contain the Planck mass. One possibility of how the Planck mass could appear such that the modifications vanish in the limit $M_{\rm P}\to \infty$ is in a ratio with the sum of mass of elementary particles 
\begin{align}\label{eq:hatXiMM}
\hat \Xi= \sum_I \frac{(a_I M_I)}{M_{\rm P}}\,,
\end{align}
where $I$ runs over all elementary particles, and $a_I$ are numerical coefficients which need to be derived or measured.

In the following, we turn to the conventional WEP-violating model to constrain the parameter $\hat \Xi$ with the most recent bounds on the E\"otv\'os parameter.

\subsection{Constraining models with MICROSCOPE data}
Before comparing this inequality with data, we first need to understand which magnitude the effect is expected to have in the macroscopic regime, in particular, because the most precise determination of the the E\"otv\'os parameter is provided by macroscopic experiments like the MICROSCOPE collaboration \cite{MICROSCOPE:2022doy}. In the following, we therefore consider the analogous relation to Eq. \eqref{eq:EotvosGen} for macroscopic bodies with masses $M_1$ and $M_2$,
\begin{equation}
	\eta\geq|\Xi_{\rm mac,1}(M_1)M_1-\Xi_{\rm mac,2}(M_2)M_2|,
\end{equation}
with the deformation parameters   $\Xi_{\rm mac,1}$ and $\Xi_{\rm mac,2}$ applying to the two macroscopic bodies independently.

At first glance, in the context of quantum gravity it appears reasonable to assume that corrections become important at the Planck scale, that is $M_{\rm P}\sim 10^{-5}\rm \,g$. For macroscopic processes, this mass scale is tiny, thus apparently leading to large deviations from existing observations. However, Planck-scale MDRs only apply to truly elementary particles, not the composites we experience in the macrocosm. Instead, it can be shown that the deformation scale for objects made up of $N$ constituents of equal type depletes by a factor $1/N$ \cite{Amelino-Camelia:2013fxa,Bosso:2023aht,Wagner:2023fmb}. In other words, the corrections scale with the mass of the microscopic constituents of the macroscopic body. We prove this scaling for the Lagrangian \eqref{eqn:NRLagkappaPoincare} and a simplified model of a rigid body in App. \ref{app:CompPart}. Thus, for composite objects of masses $M_I=N_I M_{c,I}$ each made up of one type of constituent with masses $M_{\rm c, I}$, we obtain $\Xi_{\rm mac,I} (M_{I})=\Xi_{c,I} M_{c,n}/M_n,$ where $\Xi_{c,I}$ denotes the perturbation parameter that applies to the microscopic constituent. As a result, the E\"otv\'os parameter reads
\begin{equation}\label{eq:etacomp}
	\eta\gtrsim|\Xi_{\rm c,1}(M_{\rm c,1})M_{\rm c,1}-\Xi_{\rm c,2}(M_{\rm c,2})M_{c,2}|.
\end{equation}
In other words at first order in the correction, the E\"otv\'os factor solely depends on the masses of the constituents. This implies that, barring deviations from the idealistic model of a rigid body entertained in App. \ref{app:CompPart}, for composites of \emph{equal constituents} ($\Xi_{\rm c,1}=\Xi_{\rm c,2},$ $M_{\rm c,1}=M_{\rm c,2}$) the E\"otv\'os factor generally satsifies
\begin{equation}
	\eta=0.
\end{equation}
This would be expected, for example, for rigid bodies made up of \emph{the same} constituents (neglecting defects). In short, while the violation of the WEP is sourced by a mass-dependence of the acceleration at the fundamental level, at the macroscopic level it is due to a dependence on the microscopic structure. Note that both of these types are covered by the definition of the WEP given in the introduction, and both generically lead to non-zero E\"otv\'os parameters.

However, the crude model for composite objects advertised in App. \ref{app:CompPart} does not faithfully describe atoms or, even less so, atomic nuclei. Their mass does not stem from their constituents (quarks), but the strong interactions between them. Indeed, there is no clear notion of number of constituents inside a nucleus. In the end, a precise determination of the corrections to $\eta$ can only be accounted for in a field-theoretical setup.

For the purpose of the present paper, we are content with an order-of-magnitude estimate. Therefore, we assume the constituents to be the atoms, the rigid bodies are constructed from, and divide the resulting mass difference by the order of magnitude of the involved elementary particles in those nuclei. In estimating this number, we sum the effective number of valence quarks for nucleons (\ie 3) and add the number of electrons in the atoms. In other words we assume that $\Xi_{\rm c,1}\simeq\Xi_{\rm c,2}\simeq \hat\Xi/N,$ where $\hat\Xi,$ we recall, denotes the deformation parameter applying to elementary particles. In the literature on quantum gravity phenomenology it is expected to be mass independent and of the order of the Planck length. Furthermore, $N$ accounts for the order of magnitude of the elementary particles in the involved nuclei.

At this stage, we can compare our findings to MICROSCOPE data \cite{MICROSCOPE:2022doy}, which has bounded the E\"otv\'os parameter to $\eta<10^{-15}.$ This measurement is based on comparing a mass made of platinum-rhenium alloys with a mass containing titanium-aluminium-vanadium alloys. As both contain the first named element (platinum and titanium) to 90\%, we neglect the other compounds such that the E\"otv\'os factor reads
\begin{equation}
\eta\gtrsim \frac{\hat\Xi}{N}|M_{\rm c,Pt}-M_{\rm c, Ti}|,
\end{equation}
with the mass of a Pt${}^{195}$ isotope $M_{\rm c,Pt}\simeq 195$GeV$/c^2$ and the mass of a Ti${}^{48}$ isotope $M_{\rm c,Ti}\simeq 48$GeV$/c^2$ as well as $N\simeq 10^{2}$. As a result, we obtain the experimental constraint on $\hat \Xi$
\begin{equation}\label{eq:micbounds}
	\hat\Xi^{-1}\gtrsim10^{15}\,{\rm GeV}/c^2.
\end{equation}
Being just three orders of magnitude shy from the Planck scale and considering the increased control over the process in comparison to astrophysical observations \cite{HESS:2011aa,MAGIC:2017vah,MAGIC:2020egb}, this is a very competitive bound. We stress that we have not shown, that the dynamics encoded in the post-Newtonian Finslerian free point-particle action based on Eq. \eqref{eq:fkappa} allow for a (deformed) relativity principle. Therefore, this bound, while being inspired from DSR should still be regarded as constraining a Lorentz-invariance violating scenario. 

Having demonstrated how MDRs deviate from the WEP, it is time to propose new classes of MDRs which actually satisfy the WEP and thus allow for a purely geometric description of gravity.

\subsection{General MDRs satisfying the WEP}\label{ssec:MDRsWEP}

In the previous discussion we saw that the mass scaling of the term $\Xi(M) h(x,p)$  is crucial for MDRs to satisfy the WEP. Assuming $h(x,p)$ to be $n$-homogeneous w.r.t the relativistic 4-momenta, a generic MDR satisfying the WEP is
\begin{align}
	g^{ab}(x)p_a p_b + M^{2-n} \hat \Xi h(x,p) = M^2\,.
\end{align}
To disentangle the mass dependence we can solve for $M^2$ and get a generic mass independent Hamiltonian, which leads to an MDR satisfying the WEP
\begin{align}
	M^2 = 	(g^{ab}(x)p_a p_b)+  \hat \Xi (g^{ab}(x)p_a p_b)^{\frac{2-n}{2}} h(x,p)  = H(x,p)\,.
\end{align}
Then, going through the algorithm outlined in \cite{Lobo:2020qoa} to obtain the corresponding Finsler function, we find
\begin{align}
	F(x,\dot x)
	&= \sqrt{g(\dot x, \dot x)}\left(M - \frac{\hat \Xi}{2} \frac{(g^{ab}(x)\bar p_a\bar  p_b)^{\frac{2-n}{2}}  h(x,\bar p)}{ M}\right) \\
	=& M \sqrt{g(\dot x, \dot x)}\left(1 - \frac{\hat \Xi}{2}(g^{ab}(x)\hat p_a\hat  p_b)^{\frac{2-n}{2}}  h(x,\hat p)\right)\,,\quad \hat p^a = \frac{\dot x^a}{\sqrt{g(\dot x, \dot x)}}\,,
\end{align}
which indeed leads to a spatial Lagrangian $f$ which is independent of $M$ and so is the spatial acceleration equation \eqref{eq:acceqallorders}. Thus, the WEP is valid by construction.

For general polynomial MDRs compatible with the WEP we may consider expressions of the type
\begin{align}
\boxed{
    H(x,p) = g^{ab}(x)p_a p_b + \sum_{i=0}^m \hat{\Xi}^i \left( (g^{cd}(x)p_c p_d)^{1-i} \right) h^{a_1...a_{i}} p_{a_1}...p_{a_i}\,.
    }
\end{align}
Motivated by our finding that MDRs can lead to spatial acceleration equations which contain the particle's mass we close this discussion providing the necessary condition for a spatial Lagrangian $f$ to satisfy the WEP, and arguing why the homogeneity in $M$ is the obvious solution.

Consider a spatial Lagrangian $f$ of the type
\begin{align}
    f = \sum_{i=0}^N M^i \hat f^{(i)}\,,
\end{align}
where the series coefficients $\hat f^{(i)}$ are independent of $M$. Then, the spatial acceleration equation reads
\begin{align}\label{eq:MDRWEPcheck}
    \sum_{i=0}^N M^i\left(\hat{f}^{(i)}_{\alpha\beta}a^\beta-c^2\partial_\alpha \hat{f}^{(i)}-c(\partial_t\hat{f}^{(i)}_\alpha+v^\beta\partial_\beta\hat{f}^{(i)}_\alpha)\right)=0\,.
\end{align}
If the WEP is valid, \ie $\D a^\alpha/\D M=0$ and $\D v^\alpha/\D M=0$, then
\begin{align}
    \hat{f}^{(n)}_{\alpha\beta}a^\beta=c^2\partial_\alpha \hat{f}^{(n)} c(\partial_t\hat{f}^{(n)}_\alpha+v^\beta\partial_\beta\hat{f}^{(n)}_\alpha)\,,
\end{align}
must hold for each $n$. Defining the inverses $f_{(n)}^{\alpha\beta}f^{(n)}_{\beta\gamma}=\delta^{\alpha}_\gamma,$ this implies that
\begin{equation}
    f^{\alpha\beta}_{(n)}\left(c^2\partial_\beta \hat{f}^{(n)}-c(\partial_t\hat{f}^{(n)}_\beta+v^\gamma\partial_\gamma\hat{f}^{(n)}_\beta)\right)=f^{\alpha\beta}_{(m)}\left(c^2\partial_\beta \hat{f}^{(m)}-c(\partial_t\hat{f}^{(m)}_\beta+v^\gamma\partial_\gamma\hat{f}^{(m)}_\beta)\right)\qquad \forall n,m.
\end{equation}
This is a strong constraint on the series coefficients, which is clearly solved by $f^{(n)}=f^{(m)}+\rm constant$ for all $n,$ $m.$ In other words, $f$ should be of the kind
\begin{equation}
    f=\mathcal{M}(M)\hat{f} + \rm constants,
\end{equation}
for some dimensionless function of the mass $\mathcal{M}.$ Understanding $f$ as a spatial Lagrange function as suggested by \eqref{eq:geodspat}, we find that admissible additions to $f$ can actually be boundary terms which do not modify the equations of motion. Thus, we may write
\begin{equation}
    f=\mathcal{M}(M)\hat{f} + \mathrm{boundary\ term}.
\end{equation}
Note that the function $\mathcal{M}$ amounts to a constant prefactor of a Lagrangian which has no effect whatsoever on the dynamics. Therefore, the Lagrangian can as well be rescaled such that $\mathcal{M}=1,$ implying that it is entirely independent of the mass. Clearly, if the Lagrangian is independent of the mass, so are the equations of motion, automatically satisfying the WEP.

\section{Discussion}\label{sec:Discussion}
In this article we demonstrated that large classes of MDRs on curved spacetimes lead to deviations from the WEP. The result is summarized in Theorem \ref{thm:WEPMDR}. It states that the terms in the relativistic Hamiltonian, which defines the MDR, must have a specific mass scaling property with respect to each other. This can be understood in the way that the product between the, possibly mass dependent, coupling constant $\Xi$ and the perturbation function $h$ must scale with a factor of $M^2$, or that the coupling is independent of the mass and the perturbation function must be homogeneous of degree two with respect to the momenta. In Sec. \ref{ssec:MDRsWEP} we showed that a mass dependent coupling constant in the MDR is equivallent to having a mass indepent MDR that is rescaled such that the whole dispersion relation defining Hamiltonian is homogeneous in the momenta, up to a dynamically trivial term. Inhomogeneous MDRs with a mass independent coupling function $\Xi=\hat{\Xi}$, which emerge for example in the context of quantum gravity phenomenology, deformed relativity models, non-commutative geometry, generalized uncertainty principles or Lorentz invariance violation, inevitably predict a spatial gravitational acceleration which depends on the particle’s mass.

In order to reach this insight we have discovered several intermediate results, which are significant on their own:
\begin{itemize}
    \item We derived the $3+1$ decomposition and the general slow-velocity expansion of a generic parametrization invariant relativistic point-particle action to order $c^{-2}$ in \eqref{eqn:PertLag}.
    \item In all generality, we proved for the  $3+1$ decomposed generic parametrization invariant relativistic point-particle action that the spatial acceleration \eqref{eq:acceqallorders} is obtained as Euler-Lagrange equation \eqref{eq:geodspat} of the time-coordinate parametrized relativistic Lagrangian, which we call the spatial Lagrangian $f$.
    \item We derived the nonrelativistic expansion of the spatial acceleration equation to order $c^{-2}$ in equations \eqref{eq:accnonrelsimp}, \eqref{eq:accNewt}, \eqref{eq:acc1PN} and \eqref{eq:acc2PN}.
\end{itemize}
The essential step to reach Theorem \eqref{thm:WEPMDR} consisted in applying the above findings to parametrization invariant relativistic point-particle actions induced by MDRs.

Beyond just proving Theorem \eqref{thm:WEPMDR}, we quantified the deviations from the WEP for non-homogeneous MDRs by calculating the E\"otv\'os factor for general polynomial MDRs \eqref{eq:eotMDR} and for the example of the first-order $\kappa$-Poincar\'e MDR in the bicrossproduct basis \eqref{eq:eotkappa} explicitly. Last but not least, we conjectured how MDRs can be sonstructed in order to satsify the WEP. For inhomogeneous MDRs a mass dependent coupling constant can be constructed in terms of the correct power of the particles Mass and a dimensionless, mass independent, coupling constant given for example by a ratio between elementary particle masses and the Planck mass in \eqref{eq:hatXiMM}. A  general form of $2$-homogeneous MDRs was constructed in \eqref{eq:MDRWEPcheck}, by a resealing of the deviation from GR by the metric norm of the particles four-momenta.

Our findings are based on the assumption that in the one-particle sector the dispersion relation and the resulting Hamilton equations of motion fully determine the motion of particles. General DSR theories, like the $\kappa$-Poincar\'e framework, contain additional structure, such as modified energy-momentum addition laws and deformed local symmetries of spacetime, among them modified local Lorentz transformations \cite{Lukierski:1992dt,Majid:1994cy,Kowalski-Glikman:2002iba,Amelino-Camelia:2000stu}. On flat spacetimes, studies of the compatibility of the Finsler approach and the additional DSR principles have already been done and have been positive  \cite{Morais:2023amp,Lobo:2021yem}. On generically curved spacetimes, which is the realm of our analysis, the construction of these DSR features is not completed yet. Mostly they are understood for maximally symmetric spacetimes with attempts to homogeneous and isotropic spacetime geometry \cite{Rosati:2015pga}. Thus, a future research direction is to understand if and how the additional features of DSR theories may influence the derivation of the E\"otv\'os factor, for example by the modified identification of the particle's velocity \cite{Mignemi:2019yzn}, or a modified coupling to the gravitational field. This would go beyond our assumption that the one-particle sector of the theory is purely described by the MDR.

An additional future application of  the $3+1$ decomposition and slow-velocity expansion of generic parametrization invariant relativistic point-particle action, which is the fundamental ingredient to Finsler geometry and Finsler gravity \cite{Hohmann:2019sni,Pfeifer:2019wus}, is the pN analysis of Finsler gravity and the gravitational field of kinetic gases \cite{Hohmann:2020yia}.

In short, we have opened up three major avenues for future pursuit. First, continuing the approach of \cite{Wagner:2023fmb}, by the PN expansion, we have shifted generic MDRs to the highly controllable ultra precise table-top regime (for example matter wave interferometry), allowing for a new, competitive area of phenomenology. Second, we have opened up the possibility for astrophysical precision tests of Finsler gravity, and third we pose the question if the analysis of the WEP changes in any way when one takes full DSR frameworks into account. We expect to report back on this issue soon.

\begin{acknowledgments}
	MH gratefully acknowledges the full financial support by the Estonian Research Council through the Center of Excellence TK202 ``Fundamental Universe''.
	CP acknowledges the financial support by the excellence cluster QuantumFrontiers of the German Research Foundation (Deutsche Forschungsgemeinschaft, DFG) under Germany's Excellence Strategy -- EXC-2123 QuantumFrontiers -- 390837967 and was funded by the Deutsche Forschungsgemeinschaft (DFG, German Research Foundation) - Project Number 420243324.
\end{acknowledgments}

\appendix

\section{From the Helmholtz action to the Finsler function}\label{appx:Helmholtz}
We claimed that the Helmholtz action leads, in general, to a Finslerian length measure \eqref{eq:Fins}. Here we provide a quite general proof of this statement. We start from the Helmholtz action of free point particles subject to a dispersion relation $H(x,p)=m^2$
\begin{align}
	S_H[x,p,\lambda] = \int d\tau (p_a \dot x^a - \lambda f(H(x,p),M))\,,
\end{align}
and the corresponding field equations from the variation with respect to $\lambda$, $p$ and  $x$
\begin{align}
	f(H,M) &= 0 \Leftrightarrow H = M^2\,,\\
	\dot x^a  & = \lambda \bar{\partial}^a f = \lambda \partial_H f \bar{\partial}^a H \label{eq:dx}\,,\\
	\dot p_a &  = -\lambda \partial_a f= - \lambda \partial_H f \partial_a H \,.
\end{align}
The strategy now is to determine the homogeneity $p$ with respect to $\dot x$ to show that
\begin{align}
	S_H[x,p(x,\dot x),\lambda(x,\dot x)] = \int d\tau \ p_a(x,\dot x) \dot x^a \,,
\end{align}
is the desired Finsler function that is $1$-homogeneous with respect to $\dot x$.

Starting with Eq. \eqref{eq:dx}, a necessary condition to turn the Helmholtz action into a Finslerian length element is that
\begin{align}
	y^a := \frac{\dot x^a}{\lambda \partial_H f}=\bar{\partial}^a H \,,
\end{align}
can be solved for $p_a = p_a(x,y(\dot x,\lambda))$. Next, these momenta must be inserted into the dispersion relation and then $\lambda =\lambda(x,\dot x) $ must be found such that
\begin{align}
	H(x,p(y)) = M^2\,.
\end{align}
In order for these equations to be satisfied $\lambda(x,\dot x)$ must be $1$-homogeneous with respect to $\dot x^a$, as we will see now.

Recall the Euler theorem for homogeneous function which states that for an $n$-homogeneous function $f(x)$, the following holds
\begin{align}
	x^a \partial_a f(x) = n f(x)\,.
\end{align}
 Applying a derivative operator of this kind to $H=M^2$ yields
 \begin{align}\label{eq:dxlambda}
 		0 = \dot x^m \dot{\partial}_m H(x,p(y))
 		= \bar{\partial}^q H   \frac{\partial p_q}{\partial y^a}  \dot x^m \dot{\partial}_m y^a
 		= \bar{\partial}^q H   \frac{\partial p_q}{\partial y^a} y^a  \left( 1 - \frac{\dot x^m \dot{\partial}_m \tilde \lambda}{\tilde \lambda}  \right) \,.
 \end{align}
Since we assumed \eqref{eq:dx} can be solved for $p(x,y)$ the transformation $\frac{\partial p_q}{\partial y^a}$ matrix is non-degenerate. Thus, \eqref{eq:dxlambda} implies
\begin{align}
	\dot x^m \dot{\partial}_m \tilde \lambda =\tilde \lambda\,,
\end{align}
meaning, according to Euler's Theorem, that $\tilde \lambda$ is $1$-homogeneous with respect to $\dot x$. In turn, this implies that $y$ is $0$-homogeneous with respect to $\dot x$ and so is $p$, since $p$ depends on $\dot x$ through $y$.

The algorithm we just detailed, was explicitly applied to first order modifications of the general relativistic dispersion relation of free point particles in \cite{Lobo:2020qoa}.

\section{Explicit form of pN-expansion to the acceleration equation\label{app:acceqExpPPN}}

In this appendix, we display the $c^{-1}$ and $c^{-2}$ pN-expansion of the spatial acceleration \eqref{eq:accnonrelsimp} explicitly.

\noindent The first $(c^{-1})$ relativistic correction reads
\begin{align}\label{eq:acc1PN}
    \begin{split}
        &\ord{\bar f}{0}_{\alpha\beta}\ord{\frac{d^2 x^\alpha}{d t^2}}{1} \\
        &= \left( \ord{G}{1}_\beta + \ord{\bar f}{0}_{\alpha\beta} \ord{q}{1}^{\ \kappa\alpha}\ \ord{G}{0}_\kappa \right)\\
        &=
        \partial_\beta \ord{\bar f}{3} - \partial_t \ord{\bar f}{2}_\beta + \ord{\bar f}{0}^{\beta\kappa}\ord{\bar f}{1}_{\gamma\beta}(\partial_t \ord{\bar f}{1}_\kappa - \partial_\kappa \ord{\bar f}{2} )\\
        &+ v^{\sigma_1} \left[ \partial_\beta \ord{\bar f}{2}_{\sigma_1} - \partial_{\sigma_1} \ord{\bar f}{2}_{\beta} - \partial_t \ord{\bar f}{1}_{\sigma_1\beta}  - \ord{\bar f}{0}^{\lambda\kappa} \left(\ord{\bar f}{0}_{\lambda\sigma_1\beta}(\partial_\kappa\ord{\bar f}{2} - \partial_t \ord{\bar f}{1}_\kappa ) + \ord{\bar f}{1}_{\lambda\beta}(\partial_\kappa \ord{\bar f}{1}_{\sigma_1} - \partial_{\sigma_1} \ord{\bar f}{1}_{\kappa} - \partial_t \ord{\bar f}{0}_{\sigma_1\kappa}) \right) \right]\\
        &+ v^{\sigma_1} v^{\sigma_2}\left[ \tfrac{1}{2}\partial_\beta \ord{\bar f}{1}_{\sigma_1\sigma_2} - \partial_{\sigma_1} \ord{\bar f}{1}_{\beta\sigma_2} - \tfrac{1}{2}\partial_t \ord{\bar f}{0}_{\sigma_1\sigma_2\beta} - \ord{\bar f}{0}^{\lambda\kappa} \left(\ord{\bar f}{0}_{\lambda\sigma_2\beta}(\partial_\kappa\ord{\bar f}{1}_{\sigma_1} - \partial_{\sigma_1}\ord{\bar f}{1}_{\kappa} - \partial_t \ord{\bar f}{0}_{\kappa\sigma_1} ) + \ord{\bar f}{1}_{\lambda\beta} (\tfrac{1}{2}\partial_\kappa \ord{\bar f}{0}_{\sigma_1\sigma_2} - \partial_{\sigma_1} \ord{\bar f}{0}_{\sigma_2\kappa}) \right) \right] \\
        &+ v^{\sigma_1} v^{\sigma_2} v^{\sigma_3} \left[
        \tfrac{1}{3!}\partial_\beta \ord{\bar f}{0}_{\sigma_1\sigma_2\sigma_3} - \tfrac{1}{2}\partial_{\sigma_1} \ord{\bar f}{0}_{\beta\sigma_2\sigma_3} - \ord{\bar f}{0}_{\lambda\sigma_2\beta}\ord{\bar f}{0}^{\lambda\kappa} (\tfrac{1}{2} \partial_\kappa\ord{\bar f}{0}_{\sigma_1\sigma_3} - \partial_{\sigma_1}\ord{\bar f}{0}_{\kappa\sigma_3}  ) \right]\,.
    \end{split}
\end{align}

\noindent The second $(c^{-2})$ relativistic correction, the first non-vanishing in the usual (p)pN analysis, is given by
\begin{align}\label{eq:acc2PN}
    \begin{split}
        &\ord{\bar f}{0}_{\alpha\beta}\ord{\frac{d^2 x^\alpha}{d t^2}}{2} \\
        &=\left( \ord{G}{2}_\beta + \ord{\bar f}{0}_{\alpha\beta} \ord{q}{1}^{\ \kappa\alpha}\ \ord{G}{1}_\kappa + \ord{\bar f}{0}_{\alpha\beta} \ord{q}{2}^{\ \kappa\alpha}\ \ord{G}{0}_\kappa  \right)\\
        &=
        \partial_\beta \ord{\bar f}{4} - \partial_t \ord{\bar f}{3}_\beta
        -\ord{\bar f}{0}^{\kappa\lambda}\left(\ord{\bar f}{1}_{\beta\lambda}(\partial_\kappa \ord{\bar f}{3} - \partial_t\ord{\bar f}{2}_{\kappa})
        - (\ord{\bar f}{0}^{\nu\rho}\ord{\bar f}{1}_{\nu\lambda} \ord{\bar f}{1}_{\beta\rho} - \ord{\bar f}{2}_{\beta\lambda}) (\partial_\kappa \ord{\bar f}{2} - \partial_t \ord{\bar f}{1}_{\kappa})\right) \\
        &+ v^{\sigma_1}
        \Bigg[
        \partial_\beta \ord{\bar f}{3}_{\sigma_1} - \partial_{\sigma_1} \ord{\bar f}{3}_{\beta} - \partial_t \ord{\bar f}{2}_{\beta\sigma_1} - \ord{\bar f}{0}^{\kappa\lambda}
        \Bigg(
        \ord{\bar f}{1}_{\beta\lambda} (\partial_\kappa \ord{\bar f}{2}_{\sigma_1} - \partial_{\sigma_1} \ord{\bar f}{2}_{\kappa} - \partial_t  \ord{\bar f}{1}_{\kappa\sigma_1} )
        + \ord{\bar f}{0}_{\beta\lambda\sigma_1}(\partial_\kappa \ord{\bar f}{3} - \partial_t \ord{\bar f}{2}_{\kappa} ) \\
        &- (\ord{\bar f}{0}^{\nu\rho}\ord{\bar f}{1}_{\nu\lambda} \ord{\bar f}{1}_{\beta\rho} - \ord{\bar f}{2}_{\beta\lambda}) (\partial_\kappa\ord{\bar f}{1}_{\sigma_1} - \partial_{\sigma_1}\ord{\bar f}{1}_{\kappa} - \partial_t \ord{\bar f}{0}_{\kappa\sigma_1} )
        - ( \ord{\bar f}{0}^{\nu\rho} (\ord{\bar f}{1}_{\beta\rho} \ord{\bar f}{0}_{\nu\lambda\sigma_1} + \ord{\bar f}{1}_{\nu\lambda} \ord{\bar f}{0}_{\rho\beta\sigma_1}) - \ord{\bar f}{1}_{\beta\lambda\sigma_1})(\partial_\kappa \ord{\bar f}{2} - \partial_t\ord{\bar f}{1}_{\kappa})
        \Bigg)
        \Bigg]\\
        &+ v^{\sigma_1} v^{\sigma_2}
        \Bigg[
        \tfrac{1}{2}\partial_\beta \ord{\bar f}{2}_{\sigma_1\sigma_2} - \partial_{\sigma_1} \ord{\bar f}{2}_{\beta\sigma_2} - \tfrac{1}{2}\partial_t \ord{\bar f}{1}_{\sigma_1\sigma_2\beta} - \ord{\bar f}{0}^{\lambda\kappa}
        \Bigg(
        \ord{\bar f}{1}_{\lambda\beta} (\tfrac{1}{2}\partial_\kappa \ord{\bar f}{1}_{\sigma_1\sigma_2} - \partial_{\sigma_1} \ord{\bar f}{1}_{\sigma_2\kappa} - \tfrac{1}{2}\partial_t \ord{\bar f}{0}_{\sigma_1\sigma_2})
        + \ord{\bar f}{0}_{\lambda\sigma_2\beta}(\partial_\kappa\ord{\bar f}{2}_{\sigma_1} - \partial_{\sigma_1}\ord{\bar f}{2}_{\kappa} - \partial_t \ord{\bar f}{1}_{\kappa\sigma_1} )  \\
        &- (\ord{\bar f}{0}^{\nu\rho}\ord{\bar f}{1}_{\nu\lambda} \ord{\bar f}{1}_{\beta\rho} - \ord{\bar f}{2}_{\beta\lambda}) (\tfrac{1}{2}\partial_\kappa\ord{\bar f}{0}_{\sigma_1\sigma_2} - \partial_{\sigma_1}\ord{\bar f}{0}_{\kappa\sigma_2})
        - ( \ord{\bar f}{0}^{\nu\rho} (\ord{\bar f}{1}_{\beta\rho} \ord{\bar f}{0}_{\nu\lambda\sigma_1} + \ord{\bar f}{1}_{\nu\lambda} \ord{\bar f}{0}_{\rho\beta\sigma_1}) - \ord{\bar f}{1}_{\beta\lambda\sigma_1})(\partial_\kappa \ord{\bar f}{1}_{\sigma_1} - \partial_{\sigma_2} \ord{\bar f}{1}_{\kappa} - \partial_t\ord{\bar f}{0}_{\sigma_2\kappa})\\
        &- (\ord{\bar f}{0}^{\nu\rho}\ord{\bar f}{0}_{\beta\rho\sigma_1}\ord{\bar f}{0}_{\nu\lambda\sigma_2} - \tfrac{1}{2}\ord{\bar f}{0}_{\beta\lambda\sigma_1\sigma_2})(\partial_\kappa \ord{\bar f}{2} - \partial_t \ord{\bar f}{1}_{\kappa} )
        \Bigg)
        \Bigg]\\
        &+ v^{\sigma_1} v^{\sigma_2} v^{\sigma_3}
        \Bigg[
        \tfrac{1}{3!}\partial_\beta \ord{\Bf}{1}_{\sigma_1\sigma_2\sigma_3} - \tfrac{1}{2}\partial_{\sigma_1} \ord{\Bf}{1}_{\beta\sigma_2\sigma_3} - \tfrac{1}{3!}\partial_t \ord{\Bf}{0}_{\beta\sigma_1\sigma_2\sigma_3} - \ord{\bar f}{0}^{\lambda\kappa}
        \Bigg(
        \ord{\bar f}{1}_{\lambda\beta}( \tfrac{1}{3!}\partial_\kappa \ord{\bar f}{0}_{\sigma_1\sigma_2\sigma_3} - \tfrac{1}{2} \partial_{\sigma_1} \ord{\bar f}{0}_{\sigma_2\kappa\sigma_3} )\\
        &+ \ord{\bar f}{0}_{\beta\lambda\sigma_1} (\tfrac{1}{2}\partial_\kappa \ord{\bar f}{1}_{\sigma_2\sigma_3} - \partial_{\sigma_2} \ord{\bar f}{1}_{\sigma_3\kappa} - \tfrac{1}{2}\partial_t \ord{\bar f}{0}_{\sigma_2\sigma_3\kappa}- ( \ord{\bar f}{0}^{\nu\rho} (\ord{\bar f}{1}_{\beta\rho} \ord{\bar f}{0}_{\nu\lambda\sigma_1} + \ord{\bar f}{1}_{\nu\lambda} \ord{\bar f}{0}_{\rho\beta\sigma_1}) - \ord{\bar f}{1}_{\beta\lambda\sigma_1}) ( \tfrac{1}{2}\partial_\kappa \ord{\bar f}{0}_{\sigma_2\sigma_3} - \partial_{\sigma_2} \ord{\bar f}{0}_{\kappa\sigma_3} )\\
        &- (\ord{\bar f}{0}^{\nu\rho}\ord{\bar f}{0}_{\beta\rho\sigma_1}\ord{\bar f}{0}_{\nu\lambda\sigma_2} - \tfrac{1}{2}\ord{\bar f}{0}_{\beta\lambda\sigma_1\sigma_2}) (\partial_\kappa \ord{\bar f}{1}_{\sigma_3} - \partial_{\sigma_3} \ord{\bar f}{1}_{\kappa} - \partial_t \ord{\bar f}{0}_{\kappa\sigma_3} )
        \Bigg)
        \Bigg]\\
        &+ v^{\sigma_1} v^{\sigma_2} v^{\sigma_3} v^{\sigma_4}
        \Bigg[
        \tfrac{1}{4!}\partial_\beta \ord{\Bf}{0}_{\sigma_1\sigma_2\sigma_3\sigma_4} - \tfrac{1}{3!}\partial_{\sigma_1} \ord{\Bf}{0}_{\beta\sigma_2\sigma_3\sigma_4} - \ord{\bar f}{0}^{\lambda\kappa}
        \Bigg(
        \ord{\bar f}{0}_{\beta\lambda\sigma_1} (\tfrac{1}{3!}\partial_\kappa \ord{\bar f}{0}_{\sigma_2\sigma_3\sigma_4} - \tfrac{1}{2}\partial_{\sigma_2}\ord{\bar f}{0}_{\kappa\sigma_3\sigma_4} )\\
        &- (\ord{\bar f}{0}^{\nu\rho}\ord{\bar f}{0}_{\beta\rho\sigma_1}\ord{\bar f}{0}_{\nu\lambda\sigma_2} - \tfrac{1}{2}\ord{\bar f}{0}_{\beta\lambda\sigma_1\sigma_2}) (\tfrac{1}{2}\partial_\kappa \ord{\bar f}{0}_{\sigma_3 \sigma_4} - \partial_{\sigma_3}\ord{\bar f}{0}_{\kappa \sigma_4} )
        \Bigg)
        \Bigg]\,.
    \end{split}
\end{align}

\section{Dynamics of composite particles\label{app:CompPart}}

In the following, we derive the dynamics of composite objects as a collection particles following the dynamics of a Lagrangian of the kind given in Eq. \eqref{eqn:NRLagkappaPoincare}, in order to argue for the E\"otv\'os factor in \eqref{eq:etacomp}. The Lagrangian \eqref{eqn:NRLagkappaPoincare} describes the motion of a nonrelativistic single particle in a gravitational potential with an MDR of  $\kappa$-Poincar\'e bicrossproduct type. In so doing, we make a number of simplifying assumptions to obtain an order-of-magnitude estimate of the effect of the deformation on macroscopic-particle motion:
	\begin{itemize}
		\item We assume the body to be rigid, thus neglecting relative motion of its constituents.
		\item We consider a small body such that changes of the background gravitational field as well as the deformation are negligible within its extent.
		\item We define center-of-mass (macroscopic) and relative coordinates of an $N$-particle system as
		\begin{align}
			\vec{x}_{\rm mac}=\frac{\sum_{n=1}^NM_n\vec x_n}{\sum_{n=1}^NM_n},&&\vec{x}_{\rm rel,n}=\vec{x}_{\rm mac}-\vec{x}_n,\label{eqn:COMCoord}
		\end{align}
		where $\vec x_n$ and $M_n$ stand for position and mass of the $n$th particle. While this definition may have to be modified on a quantum spacetime, it has been recently argued that such a modification would be inconsistent on $\kappa$-Minkowski \cite{Amelino-Camelia:2020vvl}. Furthermore, the authors have shown that Eq. \eqref{eqn:COMCoord} has a particularly straightforward interpretation as center-of-mass coordinates if relative velocities vanish as we assumed above.\footnote{Indeed, the algebra as well as the coalgebra of center-of-mass coordinates/momenta and relative coordinates/momenta generally mix. However, this mixing disappears in the limit of vanishing relative momenta, in which center-of-mass and relative quantities exactly satisfy a rescaled version of the $\kappa$-Poincar\'e Hopf-algebra, see \cite{Amelino-Camelia:2020vvl}.} These results go hand in hand with a resolution of the soccer-ball problem based on the idea that the total momentum of a collection of particles is not modified \cite{Amelino-Camelia:2014gga}.
	\end{itemize}
	In short, we assume the macroscopic body to be a collection of interacting particles, moving with equal velocities into the same direction, whose extent is small in comparison to background length scales.

	In conventional units, the single-particle Lagrangian given in Eq. \eqref{eqn:NRLagkappaPoincare} reads
	\begin{align}
		L_{\rm sp}=-Mf =   -Mc^2 +  \left(1 - \Xi M \right)\tfrac{M}{2}\vec{v}^2 -(1-2\Xi M)M\phi - \tfrac{\Xi M^2}{2} \left( \ord{\vec{Z}}{1}\cdot\ord{\vec{Z}}{1} + 2 \vec{v}\cdot\ord{\vec{Z}}{1}\right).\,
	\end{align}
	Neglecting the constant bit and considering a system of $N$ identical particles interacting via an unspecified potential energy $V_{nm}$, we obtain the combined Langrangian
	\begin{align}
		L_{\rm mac} =   \sum_{n=1}^N\left[\left(1 - \Xi M_n \right)\tfrac{M_n}{2}\vec{v}_n^2 -(1-2\Xi M_n)M_n\phi_n - \tfrac{\Xi M_n^2}{2} \left( \ord{\vec{Z}}{1}_n\cdot\ord{\vec{Z}}{1}_n + 2 \vec{v}_n\cdot\ord{\vec{Z}}{1}_n\right)\right]-\sum_{n\neq m}V_{nm},\,
	\end{align}
	where $\phi_n=\phi(x_n),$ $\vec{Z}_n=\vec{Z}(x_n)$ and $\vec{v}_n=\dot{\vec x}_n.$ The velocities of the single particles can then be expressed as a sum of (macroscopic) center-of-mass velocity and relative velocities $\vec v_n=\vec v_{\rm mac}+\vec v_{{\rm rel},n}.$ Neglecting relative motion (such that $\vec v_{{\rm rel},n}=0$) and the extension of the system of particles with respect to the background fields $\phi$ and $\vec Z$  (\ie $\phi(\vec x_n)\simeq\phi(\vec x_{\rm mac})$), we obtain the Lagrangian
	\begin{equation}
		L_{\rm mac} \simeq   \sum_{n=1}^N\left[\left(1 - \Xi M_n \right)\tfrac{M_n}{2}\vec{v}_{\rm mac}^2 -(1-2\Xi M_n) M_n \phi(\vec x_{\rm mac}) - \tfrac{\Xi  M_n^2}{2}
		\left( \ord{\vec{Z}}{1}\cdot\ord{\vec{Z}}{1}(x_{\rm mac})  + 2 \vec{v}_{\rm mac}\cdot\ord{\vec{Z}}{1}(x_{\rm mac}) \right)\right]-\sum_{n\neq m}V_{nm},
	\end{equation}
	where we dropped the constant interaction term. The mass of the macroscopic body amounts to $M_{\rm mac}\simeq \sum_{n=1}^NM_n$ such that we can rewrite the Lagrangian as
	\begin{equation}
		L_{\rm mac} \simeq   \left(1 - \frac{\Xi \sum_{n=1}^NM_n^2}{M_{\rm mac}} \right)\tfrac{M_{\rm mac}}{2}\vec{v}_{\rm mac}^2 -\left(1-2\frac{\Xi\sum_{n=1}^NM_n^2}{M_{\rm mac}}\right)M_{\rm mac} \phi -\tfrac{M_{\rm mac}}{2} \frac{\Xi\sum_{n=1}^NM_n^2}{M_{\rm mac}}
		\left( \ord{\vec{Z}}{1}\cdot\ord{\vec{Z}}{1} + 2 \vec{v}_{\rm mac}\cdot\ord{\vec{Z}}{1}\right)-\sum_{n\neq m}V_{nm}.
	\end{equation}
Note that for the purpose of this paper we are interested in the scaling of the terms induced by the external gravitational field and the MDR, not the internal interaction $V_{nm}$. Therefore, we concentrate of the former.

We find that corrections to composite-object motion do not scale as $\Xi M_{\rm mac}$ but instead as $\Xi \sum_{n=1}^NM_n^2/M_{\rm mac}.$ If, for example, an object is made up of only one type of constituent such that $M_n=M$ for all $n,$ we obtain
\begin{align}
	L_{\rm mac} \simeq&   \left(1 - \Xi M \right)\tfrac{M_{\rm mac}}{2}\vec{v}_{\rm mac}^2 -\left(1-2\Xi M\right)M_{\rm mac} \phi -\tfrac{M_{\rm mac}}{2} \Xi M
	\left( \ord{\vec{Z}}{1}\cdot\ord{\vec{Z}}{1} + 2 \vec{v}_{\rm mac}\cdot\ord{\vec{Z}}{1}\right)-\sum_{n\neq m}V_{nm}\\
	=&\left(1 -  \frac{\Xi M_{\rm mac}}{N} \right)\tfrac{M_{\rm mac}}{2}\vec{v}_{\rm mac}^2 -\left(1-2\frac{\Xi M_{\rm mac}}{N}\right)M_{\rm mac} \phi -\tfrac{M_{\rm mac}}{2} \frac{\Xi M_{\rm mac}}{N}
	\left( \ord{\vec{Z}}{1}\cdot\ord{\vec{Z}}{1} + 2 \vec{v}_{\rm mac}\cdot\ord{\vec{Z}}{1}\right)-\sum_{n\neq m}V_{nm}.
\end{align}
In other words, all corrections are suppressed by a factor of $1/N.$

\bibliographystyle{unsrt}
\bibliography{bib}

\end{document}